\documentclass[12pt, a4paper]{article}
\usepackage[height=25cm,width=16cm]{geometry}
\usepackage{feynmp}
\usepackage{amsmath}
\usepackage{ascmac}
\usepackage{dcolumn}
\usepackage{bm,here}
\usepackage[FIGTOPCAP]{subfigure}
\usepackage{comment}
\usepackage{ifpdf}
\usepackage{slashed}
\usepackage{colortbl}
\usepackage{color}
\usepackage{ulem}
\usepackage{comment}
\usepackage{braket}
\usepackage[mathscr]{eucal}
\usepackage[sort&compress, numbers, merge]{natbib}

\usepackage{cancel}

\ifpdf
\usepackage{graphicx}     %   use package without driver option
\usepackage[bookmarksopen,colorlinks=true,linkcolor=bblue,citecolor=ppink,urlcolor=ppink]{hyperref}
\else     % For (p)LaTeX + dvipdfmx
\usepackage[dvipdfmx]{graphicx}     %   use package with driver option
\usepackage[dvipdfmx,bookmarksopen, colorlinks=true, linkcolor=bblue, citecolor=ppink, urlcolor=ppink]{hyperref}
\fi
\usepackage{subcaption}
\usepackage{multicol}
\definecolor{red}{rgb}{1,0,0}
\definecolor{ppink}{rgb}{0.921545,0.440586,0.687243}
\definecolor{bblue}{rgb}{0.400000,0.400000,1.000000}
\usepackage[charter]{mathdesign}
\usepackage{soul}
\usepackage{wrapfig}

\newcommand{\abs}[1]{\left\vert {#1} \right\vert}

\begin{document}

%%%%%%%%%%%%%%%%%%%%%%%%%%%%%
%%%%%%%%%%% Title %%%%%%%%%%%
%%%%%%%%%%%%%%%%%%%%%%%%%%%%%
\begin{titlepage}
\begin{center}

~\vskip 1.5cm
{\Large\bf
Detecting Sterile Neutrino Dark Matter \\ [.3em]
at MeV Gamma-Ray Observatories
}

\vskip 2.0cm
{\large
Subaru Fujisawa,
Tatsuya Hayashi, \\ [.3em]
Shigeki Matsumoto,
and
Yuki Watanabe
}

\vskip 1.5cm
{\sl Kavli IPMU (WPI), UTIAS, University of Tokyo, Kashiwa, 277-8583, Japan}

\vskip 5.5cm
\begin{abstract}
\noindent
We explore the indirect detection of sterile neutrino dark matter within the gauged $U(1)_{\rm B-L}$ extension of the Standard Model, in which three right-handed neutrinos account for neutrino masses, the baryon asymmetry, and dark matter. Focusing on the MeV mass range, we investigate two decay channels: the radiative decay $N \to \nu \gamma$, which produces a monochromatic photon, and the three-body decay $N \to e^- e^+ \nu$, which leads to a 511\,keV photon signal from positronium decay. Taking the upcoming COSI mission as a case study, we show that both signals are experimentally accessible and complementary, with the 511\,keV channel extending the sensitivity reach up to ${\cal O}(100)\,\mathrm{MeV}$. We propose a novel analysis strategy in Compton data space to isolate the diffuse 511\,keV emission. Furthermore, we incorporate, for the first time, the Sommerfeld enhancement in the decay width of $N \to e^- e^+ \nu$, enabling more accurate predictions of the signal near the kinematic threshold. The combined observation of both channels would provide a distinctive and testable signature of the sterile neutrino dark matter hypothesis.
\end{abstract}

\end{center}
\end{titlepage}

%%%%%%%%%%%%%%%%%%%%%%%%%%%%%%%%
%%%%%%%%%%% Contents %%%%%%%%%%%
%%%%%%%%%%%%%%%%%%%%%%%%%%%%%%%%
\tableofcontents
\newpage
\setcounter{page}{1}

%%%%%%%%%%%%%%%%%%%%%%%%%%%%%%%%%%%%
%%%%%%%%%%% Introduction %%%%%%%%%%%
%%%%%%%%%%%%%%%%%%%%%%%%%%%%%%%%%%%%
\section{Introduction}
\label{sec: introduction}

The existence of non-zero neutrino masses and mixing, as well as dark matter in the Universe, indicates that the Standard Model (SM) must be extended. Another major challenge in particle physics is understanding the origin of the baryon-number asymmetry of the Universe. It is known that introducing two right-handed neutrinos into the SM can account for neutrino masses and mixing via the seesaw mechanism\,\cite{Minkowski:1977sc, Yanagida:1979as, PhysRevD.20.2986} and simultaneously explain the baryon asymmetry via the leptogenesis mechanism\,\cite{FUKUGITA198645,annurev:/content/journals/10.1146/annurev.nucl.55.090704.151558}. Furthermore, the addition of a third right-handed neutrino can provide a viable dark matter candidate, namely sterile neutrino dark matter\,\cite{Asaka2005}, provided its mass exceeds $\sim 1\,\mathrm{keV}$ and its interactions are sufficiently weak to ensure stability. One such extension of the SM is the $U(1)_{\rm B-L}$ model, in which the global $U(1)_{\rm B-L}$ symmetry of the SM is promoted to a local gauge symmetry\,\cite{Khalil_2008}. In this framework, a right-handed neutrino is introduced in each generation to cancel gauge anomalies, thereby naturally incorporating three right-handed neutrinos into the theory.

Since sterile neutrino dark matter feebly interacts with SM particles, it is difficult to detect via direct detection in underground laboratories or at accelerator experiments. In contrast, indirect detection through astrophysical observations offers a promising opportunity to search for dark matter signals, provided that the dark matter has a sufficiently large decay width. Among various indirect detection channels, one of particular interest is the observation of a monochromatic photon signal (in the X-ray or gamma-ray range) from dark matter decay in our Galaxy or in nearby galaxies and clusters. This signal arises from the radiative decay process $N \to \nu \gamma$, where $N$, $\nu$, and $\gamma$ denote the sterile neutrino dark matter, the neutrino, and the photon, respectively. The resulting photon has an energy equal to half the dark matter mass, $m_N/2$, and is being extensively searched for by X-ray telescopes with unprecedented sensitivity in the keV mass range\,\cite{XRISM, predehl2021}. Moreover, for dark matter masses in the GeV range, the corresponding photon signal is also being probed with high sensitivity by GeV gamma-ray observatories, such as Fermi-LAT\,\cite{FermiLAT2009}. For dark matter masses in the MeV range, the corresponding signal was previously probed by the MeV gamma-ray observatory COMPTEL, which employed a Compton camera\,\cite{schoenfelder1993comptel}. However, the sensitivity in this range was significantly lower than in the keV and GeV ranges due to limited signal detection efficiency and substantial instrumental and albedo backgrounds. This limitation is commonly referred to as the ``MeV gap'' in the sensitivity of photon signal observations\,\cite{MeVgap}.

To overcome this observational gap, various MeV gamma-ray projects have been proposed\,\cite{DeAngelis2021, Fleischhack2021, Tomsick:2021wed}. Among them, a leading project is COSI, which also employs a Compton camera and has been approved as a NASA Small Explorer (SMEX) mission\,\cite{Tomsick:2021wed, Aramaki:2022zpw, COSI2023}. Upcoming projects, including COSI, will explore the monochromatic photon signal with improved sensitivity in the MeV mass range\,\cite{Tomsick:2021wed}. However, their accessible energy range remains limited. For instance, at COSI, the detectable energy window is constrained to $200\,\mathrm{keV} < E_\gamma < 5\,\mathrm{MeV}$ due to limitations in instrument size. We therefore propose to also utilize a different monochromatic photon signal to probe dark matter. When sterile neutrino dark matter has a mass in the MeV range, it can decay into an electron-positron pair and a neutrino ($N \to e^- e^+ \nu$), where $e^\pm$ denote the electron and positron, respectively. The resulting electrons, with energies up to ${\cal O}(100)\,\mathrm{MeV}$, are efficiently decelerated during their propagation through the Galaxy and are eventually captured by ambient electrons to form positroniums.\footnote{
    Positrons emit energetic gamma rays via inverse Compton scattering and in-flight annihilation before being captured by ambient electrons. The resulting flux, depending on the initial positron energy, may be large enough to form a sufficient amount positroniums and is thus constrained by gamma-ray observations. However, it has been shown that the constraint is not severe if the initial energy is below ${\cal O}(100)$\,MeV\,\cite{Das:2025tdh}.}
These positroniums subsequently decay into two photons with an energy of $511\,\mathrm{keV}$, with a branching ratio of approximately 25\%. On the other hand, the $511\,\mathrm{keV}$ line signal has already been observed, e.g., by the hard X-ray telescope INTEGRAL. This signal predominantly originates from the Galactic bulge, and its spatial morphology is more cuspy than expected from dark matter decay. Thus, in a decaying dark matter scenario, the observed $511\,\mathrm{keV}$ signal is likely dominated by astrophysical sources, with dark matter contributing only partially. It is therefore important to search for the diffuse $511\,\mathrm{keV}$ signal arising from dark matter decay, i.e., the component originating from regions outside the bulge. Since the morphology of the dark matter signal is expected to follow, or, due to $e^+$ propagation, even exceed, the spatial extent of the underlying dark matter profile, such a signal would appear approximately as an isotropic component of the $511\,\mathrm{keV}$ emission. INTEGRAL is not well suited for observing this isotropic component, as it relies on a coded mask to determine the incoming direction of photons. In contrast, upcoming MeV gamma-ray observatories equipped with Compton cameras are capable of effectively detecting it.

In this article, taking the COSI observation as a concrete example of an upcoming MeV gamma-ray project, we study the $511\,\mathrm{keV}$ signal originating from sterile neutrino dark matter decay by proposing an appropriate region of interest (RoI) defined in the so-called Compton data space.\footnote{
   Compton data space refers to the multi-dimensional space defined by measurable quantities associated with Compton scattering events in a detector, such as the positions and energies of successive interactions. Each event defines a “Compton cone” in this space, representing possible directions of the incoming photon. By analyzing overlaps of multiple cones, one can reconstruct the most probable source direction\,\cite{Tomsick:2021wed}.}
We demonstrate that this detection strategy can extend the accessible dark matter mass range at COSI beyond that probed by the $N \to \nu \gamma$ decay channel. Moreover, we show that two monochromatic MeV gamma-ray signals can be observed when the mass of the sterile neutrino dark matter is relatively small (i.e., less than $10\,\mathrm{MeV}$): one from the direct decay $N \to \gamma \nu$, and the other being the $511\,\mathrm{keV}$ signal originating from $N \to e^- e^+ \nu$. Since the strengths of both signals are correlated, being governed by the neutrino Yukawa interaction of the lightest right-handed neutrino, observing both would serve as a smoking gun signature of sterile neutrino dark matter. Interestingly, COSI has excellent energy resolution, less than 1\% of the observed gamma-ray energy, which enables the two signals to be distinguished even if they are nearly degenerate in energy. In such a case, the strength of the latter signal, namely the decay width of the $N \to e^+ e^- \nu$ process, is affected by a non-perturbative correction known as the threshold singularity (or Sommerfeld effect). In this work, we also calculate the decay width including this effect for the first time.

This article is organized as follows. In the next section (Section\,\ref{sec: sterile neutrino dark matter}), we present our setup, along with a brief review of sterile neutrino dark matter within the framework of the U(1)$_{\rm B-L}$ model discussed above. In Section\,\ref{sec: cosmology}, we examine the cosmological aspects of sterile neutrino dark matter. After confirming that a sufficient amount of dark matter can be produced in the early universe via the freeze-in mechanism, consistent with the observed dark matter density\,\cite{DarkMatterDensity}, we summarize the constraints on dark matter decay derived from cosmological observations. In Section\,\ref{sec: detection}, we discuss the indirect detection of sterile neutrino dark matter through astrophysical observations. We first present constraints on the dark matter parameter space from various gamma-ray and 511\,keV observations, and then explore the prospects for detecting such signals in future MeV gamma-ray observatories employing a Compton camera, taking COSI as a representative example. Section\,\ref{sec: summary and future direction} summarizes our findings and outlines possible future directions to improve the analysis developed in this article. Additional details on the calculation of the decay width for the process $N \to e^- e^+ \nu$, including the effect of the threshold singularity, are provided in Appendix\,\ref{app: Sommerfeld effect}.

%%%%%%%%%%%%%%%%%%%%%%%%%%%%%%%%
%%%%%%%%%%% Scenario %%%%%%%%%%%
%%%%%%%%%%%%%%%%%%%%%%%%%%%%%%%%
\section{The sterile neutrino dark matter}
\label{sec: sterile neutrino dark matter}

The sterile neutrino dark matter considered in this article can be identified as one of the right-handed neutrinos introduced in the minimal extension of the Standard Model (SM) by gauging the U(1)$_{\rm B-L}$ symmetry\,\cite{Jenkins:1987ue, Buchmuller:1991ce, Khalil_2008, Basso:2008iv, Fayet:1980ad, Fayet:1980rr, Fayet:1989mq, Fayet:1990wx}, which is described by the following Lagrangian:
\begin{align}
    &
    \mathcal{L} =
    \mathcal{L}_{SM}
    -\frac{1}{4} Z'_{\mu\nu}\,Z^{\prime \mu\nu}
    +(D_\mu \Phi)^\dagger (D^\mu \Phi)
    +\sum_{I = 1}^3 N_I^{\dagger} i \bar\sigma^\mu D_\mu N_I
    -g_{\rm B-L}\,Z'_\mu\,J_{\rm B-L}^\mu
    \nonumber \\
    &\qquad
    -\sum_{I = 1}^3\frac{y_I}{2}
    \left[ (N_I^c)^\dagger N_I \Phi + h.c. \right]
    -\sum_{i, I = 1}^3
    \left[
        y_{iI} H^c L_i^\dagger N_I + h.c.
    \right]
    -V(H,\Phi),
    \nonumber \\
    &
    V(H,\Phi) =
    \mu_{\Phi}^2 |\Phi|^2
    +\frac{\lambda_\Phi}{4}|\Phi|^4
    +\lambda_{\Phi H} |\Phi|^2|H|^2,
    \label{eq: Original Lagrangian}
\end{align}
where ${\cal L}_{\rm SM}$ denotes the Standard Model (SM) Lagrangian, and $Z'_\mu$ represents the U(1)$_{\rm B-L}$ gauge field, with $Z'_{\mu\nu}$ being its corresponding field strength tensor. The field $\Phi$ is the Abelian "Higgs" field that spontaneously breaks the U(1)$_{\rm B-L}$ symmetry, while $N_I$ are the right-handed neutrinos introduced to cancel anomalies in the model. The fields $H$ and $L_i$ represent the Higgs and lepton doublets of the SM, respectively. The covariant derivative acting on $\Phi$ and $N_I$ is given by $D_\mu = \partial_\mu + i q_{\Phi,\,N_I}\,g_{\rm B-L}\,Z'_\mu$, where $g_{\rm B-L}$ and $q_{\Phi,\,N_I}$ denote the U(1)$_{\rm B-L}$ gauge coupling and the corresponding charges. Here, the U(1)$_{\rm B-L}$ charges of $\Phi$ and $N_I$ are taken to be $q_\Phi = 2$ and $q_{N_I} = -1$. The U(1)$_{\rm B-L}$ current for the SM particles is given by
\begin{align}
    J_{\rm B-L}^\mu =
    \sum_f q_f\,\bar{f} \gamma^\mu f,
\end{align}
where $f = Q_i$, $U_i$, $D_i$, $L_i$, and $E_i$ denote the quark doublet, up-type quark singlet, down-type quark singlet, lepton doublet, and charged lepton singlet fields, respectively. The U(1)$_{\rm B-L}$ charges of the SM fermions are given by $q_{Q_i} = q_{U_i} = q_{D_i} = 1/3$ and $q_{L_i} = q_{E_i} = -1$. The Higgs boson does not carry a U(1)$_{\rm B-L}$ charge. The scalar potential in this model consists of $V(\Phi, H)$ in Eq.\,(\ref{eq: Original Lagrangian}) and the Higgs boson potential in ${\cal L}_{\rm SM}$, $V(H) = \mu_H^2 |H|^2 + (\lambda_H/2) |H|^4$. The Yukawa coupling between the right-handed neutrinos, $y_I$, can be real and positive without loss of generality, while the coupling between $N_I$ and $L_i$, $y_{iI}$, is generally complex.

The U(1)$_{\rm B-L}$ gauge symmetry is postulated to be spontaneously broken, assuming $\mu_\Phi^2 < 0$. The Abelian Higgs field $\Phi$ acquires a vacuum expectation value of $v_\Phi = 2\,(- \mu_\Phi^2/\lambda_\Phi)^{1/2}$, and the new particles obtain masses given by $m_{N_I} = y_I v_\Phi/\sqrt{2}$, $m_\phi = (-2\mu_\Phi^2)^{1/2}$, and $m_{Z'} = 2 g_{\rm B-L} v_\Phi$, respectively, where the field $\phi$ represents the physical mode of $\Phi$. We consider a scenario in which all the new particles, except for the lightest right-handed neutrino $N_1$, are much heavier than the scale of the electroweak symmetry breaking. After integrating out the new fields $N_2$, $N_3$, $\phi$, and $Z'$ from the Lagrangian\,(\ref{eq: Original Lagrangian}), we obtain the effective Lagrangian that describes the physics below the scale of the new particle masses as follows:
\begin{align}
    {\cal L}_{\rm eff} =
    &
    {\cal L}_{\rm SM}
    + N_1^{\dagger} i \bar\sigma^\mu \partial_\mu N_1
    -\left[ \frac{m_N}{2} (N_1^c)^\dagger N_1 + h.c.\right]
    -\sum_{i = 1}^3\,\left[y_{i1} H^c L_i^\dagger N_1 + h.c.\right]
    \nonumber \\
    &
    -\frac{g_{\rm B-L}^2}{2m_{Z'}^2}
    \left( J_{\rm B-L} - N_1^{\dagger} \bar\sigma N_1 \right)
    \cdot
    \left( J_{\rm B-L} - N_1^{\dagger} \bar\sigma N_1 \right)
    +\sum_{I = 2,3} 
    \left[
        \frac{y_{iI} y_{jI}}{2m_{N_I}}
        (H^c L_i^\dagger) (L_j^c H^c) + h.c. \right]
    \nonumber \\
    &
    +\frac{\lambda_{\Phi H} m_N}{2m_\phi^2}
    \left[
        (N_1^c)^\dagger N_1 + N_1^\dagger N_1^c
    \right] |H|^2
    +\frac{m_N^2}{8m_\phi^2 v_\Phi^2}
    \left[
        (N_1^c)^\dagger N_1 + N_1^\dagger N_1^c
    \right]^2
    + \cdots,
    \label{eq: Effective Lagrangian}
\end{align}
with $m_N \equiv y_1 v_\Phi/\sqrt{2}$ being the mass of the sterile neutrino dark matter $N_1$. We provide only the leading-order interactions explicitly obtained by integrating out each new particle. The first term in the second line corresponds to the interaction derived from the $Z'$ integration, describing current-current interactions among the dark matter and the SM particles. The next term in the same line represents the so-called Weinberg operators, derived by integrating out the right-handed neutrino fields $N_2$ and $N_3$, which account for the tiny masses of active neutrinos. The terms in the last line arise from integrating out the $\phi$ field, describing the interaction between $N_1$ and the SM Higgs boson, as well as the self-interaction of $N_1$.

First, as seen in the effective Lagrangian, the decay of the sterile neutrino dark matter $N_1$ is governed by the Yukawa interactions, $y_{i1} H^c L_i^\dagger N_1 + h.c.$ We are particularly interested in MeV-scale dark matter with a decay width comparable to the sensitivity of indirect dark matter detection, which corresponds to couplings of $y_{i1} = {\cal O}(10^{-14})$. Next, the current-current interactions and the interaction between $N_1$ and $H$ are relevant to dark matter production in the early universe. The former dominates the latter when $m_{Z'} \sim m_\phi$, $g_{\rm B-L}^2 \sim \lambda_{\Phi H}$, and $N_1$ is produced relativistically. Third, the self-scattering of dark matter predicted by the current-current interactions and the last term is suppressed by the high-energy scales $m_{Z'}$ and $m_\phi$, and thus does not play any significant role in either the present or early universe. Finally, as mentioned above, the Weinberg operator generates tiny masses for the active neutrinos. Further integrating out the dark matter field $N_1$ from the Lagrangian,(\ref{eq: Effective Lagrangian}) yields additional Weinberg operators, $y_{i1} y_{j1}\,(H^c L_i^\dagger) (L_j^c H^c)/(2 m_N) + h.c.$ These operators also contribute to the neutrino masses; however, their contribution is negligibly small compared to those arising from the integration of $N_2$ and $N_3$, as $y_{i1}$ is highly suppressed, as mentioned above.

After the electroweak symmetry is spontaneously broken, the Weinberg operators in the effective Lagrangian\,(\ref{eq: Effective Lagrangian}) lead to the (Majorana) mass matrix for the active neutrinos, expressed as $\sum_{I = 2,3} [y_{iI} y_{jI} v_{\rm EW}^2/(4m_{N_I})]\,\nu_i^\dagger \nu_j^c + h.c.$, where $v_{\rm EW} \simeq 246$\,GeV represents the vacuum expectation value of $H$. Consequently, the mass matrix can be diagonalized as
\begin{align}
    &
    {\cal L}_{\rm eff} \supset
    -\frac{1}{2}
    \nu_i^\dagger (M_\nu^*)_{ij}
    \nu_j^c + h.c.
    =
    -\frac{1}{2}
    \sum_{i = 1}^3 m_i
    (\tilde{\nu}_i^\dagger \tilde{\nu}_i^c + h.c.),
    \nonumber \\
    &
    \nu_i = (U_{\rm PMNS})_{ij} \tilde{\nu}_j,
    \,\,
    U_{\rm PMNS}^T M_\nu U_{\rm PMNS} =
    {\rm diag}(m_1, m_2, m_3),
\end{align}
where $(M_\nu)_{ij} \equiv -\sum_{I = 2,3} [y_{iI}^* y_{jI}^* v_{\rm EW}^2/(2m_{N_I})]$, $m_i$ are the masses of the active neutrinos, and the matrix $U_{\rm PMNS}$ is the so-called PMNS matrix, respectively. It should be noted that the mass matrix $M_\nu$ originates from the Yukawa interactions involving only two, rather than three, right-handed neutrinos (i.e., $N_2$ and $N_3$). As a result, the lightest active neutrino is predicted to be massless, as discussed in Ref.\,\cite{Asaka2005masses}. This is a key prediction of the sterile neutrino dark matter scenario within the U(1)$_{\rm B-L}$ extension of the Standard Model (SM).\footnote{
    The other two neutrino masses and the PMNS matrix (described by 6 parameters) can be taken arbitrarily.}

Furthermore, the electroweak symmetry breaking generates bilinear terms between the sterile neutrino dark matter $N_1$ and the active neutrinos, $[y_{i1} v_{\rm EW}/\sqrt{2}]\,\nu_i^\dagger N_1 + h.c.$, through the Yukawa interactions of $N_1$. By treating the bilinear terms perturbatively, the mass matrix involving the dark matter and the three active neutrinos is diagonalized as follows:
\begin{align}
    {\cal L}_{\rm eff} &\supset
    -\frac{1}{2} \sum_{i = 1}^3
    \left(
        m_i \tilde{\nu}_i^\dagger \tilde{\nu}_i^c
        +\sqrt{2} v_{\rm EW} y_{i1} \tilde{\nu}_j^\dagger (U_{\rm PMNS}^\dagger)_{ji} N_1
        + h.c.
    \right)
    -\frac{1}{2}
    \left(
        m_N N_1^\dagger N_1^c + h.c.
    \right)
    \\
    &\simeq
    -\frac{1}{2}
    \left(
        \sum_{i = 1}^3
        m_i \tilde{\tilde{\nu}}_i^\dagger \tilde{\tilde{\nu}}_i^c
        +
        m_N \tilde{N}_1^\dagger \tilde{N}_1^c
        + h.c.
    \right),
    \quad
    \begin{pmatrix}
        \tilde{\nu}_i
        \\
        N_1^c
    \end{pmatrix}
    \simeq
    \begin{pmatrix}
        1 & (U_{\rm PMNS}^\dagger \Theta)_{i1}
        \\
        -(\Theta^\dagger U_{\rm PMNS})_{1i} & 1
    \end{pmatrix}
    \begin{pmatrix}
        \tilde{\tilde{\nu}}_i
        \\
        \tilde{N}_1^c
    \end{pmatrix},
    \nonumber
\end{align}
where off-diagonal elements of the diagonalization matrix are $(U_{\rm PMNS}^\dagger \Theta)_{i1} = (U_{\rm PMNS}^\dagger)_{ik} \Theta_{k1}$ and $(\Theta^\dagger U_{\rm PMNS})_{1i} = (\Theta^*)_{k1} (U_{\rm PMNS})_{ki}$, with $\Theta_{i1} \equiv y_{i1} v_{\rm EW}/(\sqrt{2}\,m_N)$. Although the active neutrino masses undergo corrections of ${\cal O}(y_{i1}^2 v_{\rm EW}^2/m_N)$, these are negligibly small compared to those arising from the contributions of the other right-handed neutrinos, $N_{2,3}$. After diagonalization into the mass eigenstates, the Lagrangian\,(\ref{eq: Effective Lagrangian}) predicts the interactions for $\tilde{N}_1$:
\begin{align}
    \mathcal{L}_{\rm eff} \supset
    &
    -\left(
        \frac{g_Z}{2} Z_\mu \nu_i^\dagger \sigma^\mu \Theta_{i1} \tilde{N}_1^c
        +\frac{g}{\sqrt2} W_\mu^\dagger \ell_i^\dagger \sigma^\mu \Theta_{i1} \tilde{N}_1^c
        +\frac{y_{i1}}{\sqrt{2}} h \nu_i^\dagger \tilde{N}_1
        + h.c.
    \right)
    \label{eq: interactions 1}
    \\
    &
    +\frac{g_{\rm B-L}^2}{m_{Z'}^2} J_{\rm B-L} \cdot \tilde{N}_1^{\dagger} \bar\sigma \tilde{N}_1
    +\frac{\lambda_{\Phi H} m_N}{4m_\phi^2}
    \left[
        (\tilde{N}_1^c)^\dagger \tilde{N}_1 + h.c.
    \right] (2v_{\rm EW}\,h + h^2)
    + \cdots,
    \nonumber
\end{align}
with $Z_\mu$, $W_\mu$, $h$, and $\ell_i$ denoting the Z boson, W boson, Higgs boson, and left-handed charged lepton, respectively, while $g_Z \equiv g/\cos \theta_W$, with $g$ and $\theta_W$ being the SM SU(2)$_L$ gauge coupling and the weak mixing angle. We neglect the effects of active neutrino masses in the Lagrangian\,(\ref{eq: interactions 1}), as they are not relevant to the primary physics described by the Lagrangian, such as dark matter production in the early universe. Furthermore, we explicitly present the interactions originating from the Yukawa coupling of $N_1$ at ${\cal O}(y_{i1})$ and those from the heavy particles $Z'$ and $\phi$ at ${\cal O}(y_{i1}^0)$, while omitting $\tilde{N}_1$ self-interactions for simplicity. The mixing parameter $\Theta_{i1}$ (i.e., $y_{i1}$) can be real and positive without loss of generality by appropriately redefining the neutrino fields and the left- and right-handed charged lepton fields.

On the other hand, to discuss the decay of the dark matter $\tilde{N}_1$, whose mass lies in the MeV scale, it is convenient to employ the effective Lagrangian at the MeV scale. This Lagrangian is derived by integrating out the weak-scale SM particles from the Lagrangian (\ref{eq: interactions 1}) as
\begin{align}
    \mathcal{L}_{\rm MeV} \supset
    &
    -\sqrt2 G_F (\nu_1^\dagger \sigma^\mu \Theta_{11} \tilde{N}_1^c)
    \left[
        (2 s_W^2 + 1) (e_L^\dagger \sigma_\mu e_L)
        + 2 s_W^2 (e_R^\dagger \bar\sigma_\mu e_R)
        + \Sigma_i\,(\nu_i^\dagger \sigma_\mu \nu_i)
    \right]
    \nonumber \\
    &
    -\sqrt2 G_F (\nu_2^\dagger \sigma^\mu \Theta_{21} \tilde{N}_1^c)
    \left[
        (2 s_W^2 - 1) (e_L^\dagger \sigma_\mu e_L)
        + 2 s_W^2 (e_R^\dagger \bar\sigma_\mu e_R)
        + \Sigma_i\,(\nu_i^\dagger \sigma_\mu \nu_i)
    \right]
    \nonumber \\
    &
    -\sqrt2 G_F (\nu_3^\dagger \sigma^\mu \Theta_{31} \tilde{N}_1^c)
    \left[
        (2 s_W^2 - 1) (e_L^\dagger \sigma_\mu e_L)
        + 2 s_W^2 (e_R^\dagger \bar\sigma_\mu e_R)
        + \Sigma_i\,(\nu_i^\dagger \sigma_\mu \nu_i)
    \right]
    + h.c.,
    \label{eq: interactions 2}
\end{align}
where $G_F$ is the Fermi coupling constant, defined as $G_F = \sqrt{2}\,g^2/(8 m_W^2)$, with $m_W$ being the $W$ boson mass, $s_W \equiv \sin \theta_W$, and $e_L$ and $e_R$ denoting the left- and right-handed electron fields, respectively. We omit other interactions that are irrelevant to the dark matter decay.

Using four-component spinor fields defined as $N \equiv (\tilde{N}_1, \tilde{N}_1^c)^T$, $\nu_a \equiv (0, \nu_i)^T$ (where $a = e, \mu, \tau$), and $f \equiv (f_R, f_L)$ (with $f_L$ and $f_R$ denoting the left- and right-handed components of other SM leptons and quarks), the effective Lagrangian\,(\ref{eq: interactions 1}) can be expressed as follows:
\begin{align}
    \mathcal{L}_{\rm eff} \supset
    &
    -\frac{g_Z}{2} \Theta_{a1}
    \left(
        \bar{\nu}_a \slashed{Z} N
         +\bar{N} \slashed{Z} \nu_a
    \right)
    -\frac{g}{\sqrt2} \Theta_{a1}
    \left(
        \bar{\ell}_a \slashed{W}^\dagger P_L N
        +\bar{N} \slashed{W} P_L \ell_a
    \right)
    -\frac{y_{a1}}{\sqrt2} h
    \left(
        \bar{\nu}_a N
        +\bar{N} \nu_a
    \right)
    \nonumber \\
    &
    +\frac{g_Z^2}{2 m_{Z'}^2} J_{\rm B-L}^\mu\, \bar N \gamma_\mu \gamma^5 N
    +\frac{\lambda_{\Phi H} m_N}{4 m_\phi^2} \bar N N (2v_{\rm EW} h + h^2) + \cdots,
    \label{eq: interactions 1 II}
\end{align}
where $\slashed{Z} \equiv Z_\mu \gamma^\mu$, $\slashed{W} \equiv W_\mu \gamma^\mu$, and $P_L \equiv (1 - \gamma_5)/2$ being the projection operator onto the left-handed component. Using the same spinors, the effective Lagrangian in eq.\,(\ref{eq: interactions 2}) is
\begin{align}
    \mathcal{L}_{\rm MeV} \supset
    &
    -\sqrt2 G_F \Theta_{11}
    (\bar{\nu}_e \gamma^\mu N) 
    \left[
        (2s_W^2 + 1/2) (\bar{e} \gamma_\mu e)
        -(1/2) \bar{e} \gamma_\mu \gamma^5 e
        +\Sigma_a \left(\bar\nu_a \gamma_\mu \nu_a \right)
    \right]
    \label{eq: interactions 2 II}
    \\
    &
    -\sqrt2 G_F \Theta_{21}
    (\bar{\nu}_\mu \gamma^\mu N)
    \left[
        (2s_W^2 - 1/2) (\bar{e} \gamma_\mu e)
        +(1/2) \bar{e} \gamma_\mu \gamma^5 e
        + \Sigma_a (\bar\nu_a \gamma_\mu \nu_a)
    \right]
    \nonumber \\
    &
    -\sqrt2 G_F \Theta_{31}
    (\bar{\nu}_\tau \gamma^\mu N)
    \left[
        (2s_W^2 - 1/2) (\bar{e} \gamma_\mu e)
        +(1/2) \bar{e} \gamma_\mu \gamma^5 e
        +\Sigma_a (\bar{\nu}_a \gamma_\mu \nu_a)
    \right]
    + h.c.
    \nonumber
\end{align}

%%%%%%%%%%%%%%%%%%%%%%%%%%%%%%%%%
%%%%%%%%%%% Cosmology %%%%%%%%%%%
%%%%%%%%%%%%%%%%%%%%%%%%%%%%%%%%%
\section{Cosmology of the sterile neutrino dark matter}
\label{sec: cosmology}

In this section, we discuss the cosmology of sterile neutrino dark matter, focusing on a possible production mechanism in the early universe that accounts for the observed dark matter abundance and on the constraints imposed by cosmological observations.

First, it is crucial to elucidate the mechanisms by which a sufficient amount of dark matter could have been produced in the early universe. One such production mechanism within our framework (discussed in the previous section) is the freeze-in mechanism, wherein the annihilation and decay of the SM particles generate sterile neutrino dark matter via feeble interactions. Based on the Lagrangian\,(\ref{eq: Effective Lagrangian}), and assuming that the temperature of the universe is significantly lower than the masses of the new heavy particles $Z'$ and $\phi$, yet still higher than the electroweak scale, the possible production processes are as follows:
\begin{align}
    &
    \sigma v ( f \bar{f} \to N N ) \simeq
    \frac{q_f^2 g_{\rm B-L}^4}{12 \pi m_{Z'}^4} s,
    \quad
    \sigma v \left( H H^* \rightarrow N N \right) \simeq
    \frac{\lambda_{\Phi H}^2 m_N^2}{64\pi m_\phi^4},
\end{align}
where $f$ represents a SM fermion, and $s$ denotes the center-of-mass energy at annihilation. We do not consider processes that rely on the coupling $y_{i1}$, as they are suppressed due to its smallness. Consequently, the dominant process for producing the sterile neutrino dark matter in the early universe is pair creation through annihilations among SM particles mediated by the heavy U(1)$_{\rm B-L}$ gauge boson $Z'$. In this case, the so-called UV freeze-in mechanism is expected to be responsible for generating a sufficient amount of dark matter that is observed today. Assuming a standard inflationary scenario driven by a single inflaton field, and that the hot Big Bang universe is realized through the decay of the inflaton after inflation, the thermal history of the universe is described by the following equations\,\cite{Giudice:2001, Bernal:2025}:
\begin{align}
    &
    \frac{d\rho_\phi}{dt} =
    -3H \rho_\phi - \Gamma_\phi \rho_\phi,
    \quad
    \frac{dn_N}{dt} = -3Hn_N + 2\sum_f \braket{\sigma v (f \bar{f} \to NN)} n_f n_{\bar{f}},
    \nonumber \\
    &
    \frac{d\rho_R}{dt} =
    -4H\rho_R + \Gamma_\phi \rho_\phi - \sum_f \braket{\sigma v (f\bar{f} \to NN)} 2E_N n_f n_{\bar{f}}.
\end{align}
Here, $\rho_\phi$ and $\rho_R = g_* (\pi^2/30)\,T^4$ (where $g_*$ denotes the effective number of relativistic degrees of freedom) represent the energy densities of the inflaton and the radiation composed of SM particles, respectively. The Hubble parameter and the decay width of the inflaton are denoted by $H$ and $\Gamma_\phi$, respectively. On the other hand, the number densities of sterile neutrino dark matter and SM fermions are denoted by $n_N$ and $n_f = n_{\bar{f}} = g_f [3\zeta(3)/(4\pi^2)]\,T^3$, respectively, where $g_f$ is the number of internal degrees of freedom of the fermion $f$. Meanwhile, the thermally averaged cross-section for sterile neutrino production is given by
\begin{align}
    &
    \braket{\sigma v (f \bar{f} \to NN)} =
    \frac{1}{n_f\,n_{\bar{f}}} \int \frac{d^3p}{(2\pi)^3} \frac{d^3p'}{(2\pi)^3}
    \left(
        \frac{q_f^2 g_{\rm B-L}^4}{12 \pi m_{Z'}^4} s
    \right)
    f_f({\vec{p}})\,f_{\bar{f}}(\vec{p}') \simeq
    3.8 \times 10^{-2} \frac{q_f^2 T^2}{v_\phi^4}
\end{align}
where $v_\Phi = m_{Z'}/(2 g_{\rm B-L})$, and $f_f(\vec{p}) = f_{\bar{f}}(\vec{p}) = g_f/[\exp(|\vec{p}|/T) + 1]$. The typical energy of the sterile neutrino dark matter produced via this process is estimated as $E_N \simeq \rho_f/n_f \simeq 3T$.

By solving the above three differential equations together with the Friedmann equation,
\begin{align}
    H^2 =
    \left(\frac{\dot{a}}{a}\right)^2 \simeq
    \frac{8\pi G_N}{3}(\rho_\phi + \rho_R),
\end{align}
where $\dot{a} = da/dt$ and $G_N$ denotes Newton's constant, the dark matter abundance $\Omega_{\rm DM} h^2$, i.e., the present-day average mass density of dark matter, can be obtained as follows\,\cite{Bernal:2025}:
\begin{align}
    \Omega_{\rm DM}\,h^2 \simeq
    0.11 \times
    \left( \frac{m_N}{1\,\rm MeV} \right)
    \left( \frac{T_{\rm RH}}{10^{14}\,\rm GeV} \right)^3
    \left( \frac{10^{16}\,\rm GeV}{v_\Phi} \right)^4,
    \label{eq: abundance}
\end{align}
where $h \simeq 0.7$ is the normalized Hubble constant. Here, the reheating temperature $T_{\rm RH}$ is defined by the condition $\rho_\phi(T_{\rm RH}) = \rho_R(T_{\rm RH})$ and is related to the decay width of the inflaton by $\Gamma_\phi = [\pi^2 g_*(T_{\rm RH})/90]^{1/2} T_{\rm RH}^2 / m_{\rm pl}$, where $m_{\rm pl} = (8\pi G_N)^{-1/2}$ is the reduced Planck mass. The initial conditions used to solve the above equations and obtain the result given in Eq.\,(\ref{eq: abundance}) are $\rho_\phi(t_I) = 3 m_{\rm pl}^2 H_I^2$ and $\rho_R(t_I) = n_N(t_I) = 0$, where $H_I$ is the Hubble parameter at the time $t = t_I$. Note that the production of sterile neutrino dark matter at $T(t) = T_{\rm RH}$ dominates the dark matter abundance, so the final abundance does not depend on the value of $H_I$. It follows from the result in Eq.\,(\ref{eq: abundance}) that sterile neutrino dark matter is produced with the correct relic abundance observed today when the reheating temperature $T_{\rm RH}$ and the scale of new heavy particles $v_\Phi$ satisfy a certain relation, which is also consistent with the condition required for successful seesaw mechanism and thermal leptogenesis.

Next, we examine how the decay of sterile neutrino dark matter in the early universe impacts the expansion and thermal history of the universe, as constrained by BBN and CMB observations. According to the Lagrangian\,(\ref{eq: interactions 2 II}), the dark matter is shown to decay into three active neutrinos, with the partial decay width (summed over all flavors) being\,\cite{Atre:2009rg}\,\footnote{
    The complete set of possible final states for the decay is $\nu_e \nu_e \bar{\nu}_e$, $\bar{\nu}_e \nu_e \bar{\nu}_e$, $\nu_e \nu_\mu \bar{\nu}_\mu$, $\bar{\nu}_e \nu_\mu \bar{\nu}_\mu$, $\nu_e \nu_\tau \bar{\nu}_\tau$, $\bar{\nu}_e \nu_\tau \bar{\nu}_\tau$, $\nu_\mu \nu_e \bar{\nu}_e$, $\bar{\nu}_\mu \nu_e \bar{\nu}_e$, $\nu_\mu \nu_\mu \bar{\nu}_\mu$, $\bar{\nu}_\mu \nu_\mu \bar{\nu}_\mu$, $\nu_\mu \nu_\tau \bar{\nu}_\tau$, $\bar{\nu}_\mu \nu_\tau \bar{\nu}_\tau$, $\nu_\tau \nu_e \bar{\nu}_e$, $\bar{\nu}_\tau \nu_e \bar{\nu}_e$, $\nu_\tau \nu_\mu \bar{\nu}_\mu$, $\bar{\nu}_\tau \nu_\mu \bar{\nu}_\mu$, $\nu_\tau \nu_\tau \bar{\nu}_\tau$, and $\bar{\nu}_\tau \nu_\tau \bar{\nu}_\tau$.}
\begin{equation}
    \Gamma[N \to \nu\nu\nu] =
    \frac{G_F^2 m_N^5}{96\pi^3}
    \sum_{i=1}^3 \Theta_{i1}^2.
    \label{eq: 3nu}
\end{equation}
This decay provides the leading contribution to the total decay width of the sterile neutrino dark matter when its mass is ${\cal O}(\mathrm{MeV})$, potentially affecting the expansion history of the universe. When a substantial fraction of dark matter decays, the energy budget of the universe is altered, leading to a modification of the Hubble parameter and, consequently, the early universe cosmology. Precise CMB observations and baryon acoustic oscillation measurements disfavor cosmologies with such contributions from new physics, thereby providing an upper bound on the width\,(\ref{eq: 3nu}): $\Gamma[N \to \nu \nu \nu] < 1.3 \times 10^{-19}\,\mathrm{s}^{-1}$ at 95\% C.L.\,\cite{Alvi_2022}, assuming that it is the dominant decay process. This constraint can be translated into a bound on the mixing angle: $\Theta^2 < 6.2 \times 10^{-16}\,(1\,\mathrm{MeV}/m_N)^5$, under the assumption $\Theta_{11} = \Theta_{21} = \Theta_{31} \equiv \Theta$.

In addition to the above decay into three active neutrinos, the dark matter can also decay into an active neutrino and an electron-positron pair, provided the energy is above the $e^- e^+$ threshold. The partial decay width, summed over all neutrino flavors, is\,\cite{Atre:2009rg}:
\begin{align}
    &
    \Gamma[N \rightarrow e^- e^+ \nu]
    = \frac{G_F^2 m_N^5}{192 \pi^3}
    \sum_{i=1}^3 \Theta_{i1}^2
    \left[
        (2 s_W^2 + \delta_{i1} - 1/2)^2\, F_1\left(\frac{m_e}{m_N}\right)
        + \frac{1}{4}\,F_2\left(\frac{m_e}{m_N}\right)
    \right],
    \label{eq: eenu} \\
    &
    F_1(x) \equiv
    (1-10x^2+18x^4-36x^6) \sqrt{1-4x^2}
    + 48x^6 (2-3x^2)
    \log\left[\frac{1+\sqrt{1-4x^2}}{2x} \right],
    \nonumber \\
    &
    F_2(x) \equiv
    (1-18x^2-22x^4+12x^6) \sqrt{1-4x^2}
    + 48x^4 (2-2x^2+x^4) \log\left[\frac{1+\sqrt{1-4x^2}}{2x} \right].
    \nonumber
\end{align}
This decay width is based on a perturbative calculation, which is generally invalid near the threshold region, where a long-range force, namely, the Coulomb force, acts between the final-state $e^- e^+$, leading to the so-called threshold singularity, known as the Sommerfeld effect. The decay amplitude receives corrections from $n$-loop diagrams, with ${\cal O}(e\alpha^n)$ contributions (where $\alpha = e^2/(4\pi)$ is the fine-structure constant), accompanied by an additional kinematical factor $[m_e/(m_N - 2m_e)]^n$ arising from the infrared regions of the loop momenta. This factor enhances the amplitude at each order in perturbation theory, resulting in the breakdown of the naive perturbative calculation in the threshold region. To incorporate this effect, the decay width is computed using the so-called potential non-relativistic Lagrangian method, as described in Appendix\,\ref{app: Sommerfeld effect}, and the final expression is given by
\begin{align}
    \Gamma^{(S)}[N \to e^- e^+ \nu] =
    \Gamma[N \rightarrow e^- e^+ \nu]
    +
    \Delta \Gamma[N \to (e^-e^+)_{{}^1S_0} \nu]
    +
    \Delta \Gamma[N \to (e^-e^+)_{{}^3S_1} \nu],
    \label{eq: Sommerfeld eenu}
\end{align}
where the corrections to the decay widths (i.e., $\Delta \Gamma$'s) on the right-hand side are defined as
{\small
\begin{align}
    &
    \frac{\Delta \Gamma[N \to (e^-e^+)_{{}^1S_0} \nu]}
    {G_F^2 m_N^5/(32\pi^3)}
    =
    \sum_a \Theta_{a1}^2\,I(x_N),
    ~~
    \frac{\Delta\Gamma[N \rightarrow (e^-e^+)_{{}^3S_1} \nu]}
    {G_F^2 m_N^5/(32\pi^3)}
    =
    3\sum_a \Theta_{a1}^2 \left(4s_W^2 + 2\delta_{a1} -1\right)^2\,I(x_N),
    \nonumber \\
    &\qquad\qquad\qquad\quad~~
    I(x) =
    \int_0^{x-2} dy\,
    \left(
        \frac{\pi\alpha/\sqrt{y}}{1 - e^{-\pi \alpha/\sqrt{y}}} -1
    \right)
    \frac{\sqrt{y} (\sqrt{x - 1 - y} - 1)^2}{2 \sqrt{x - 1 - y}},
    \label{eq: Decay with Sommerfeld}
\end{align}
}\noindent
where $x_N = m_N/m_e$. The corrections $\Delta \Gamma$ become negligibly small compared to the tree-level contribution, $\Gamma[N \rightarrow e^- e^+ \nu]$, when $m_N \gg m_e$. As expected, the corrections vanish when the long-range force is turned off (i.e., $\alpha \to 0$), due to the factor $(\pi \alpha/\sqrt{y})/[1 - \exp(-\pi \alpha/\sqrt{y})] - 1$. The correction increases the decay width by approximately 5\,\% (20\,\%, 70\,\%) when $m_N$ is nearly degenerate with $2m_e$ at the 10\,\% (1\,\%, 0.1\,\%) level.

\begin{figure}[t]
    \centering
    \includegraphics[keepaspectratio, scale=0.5]{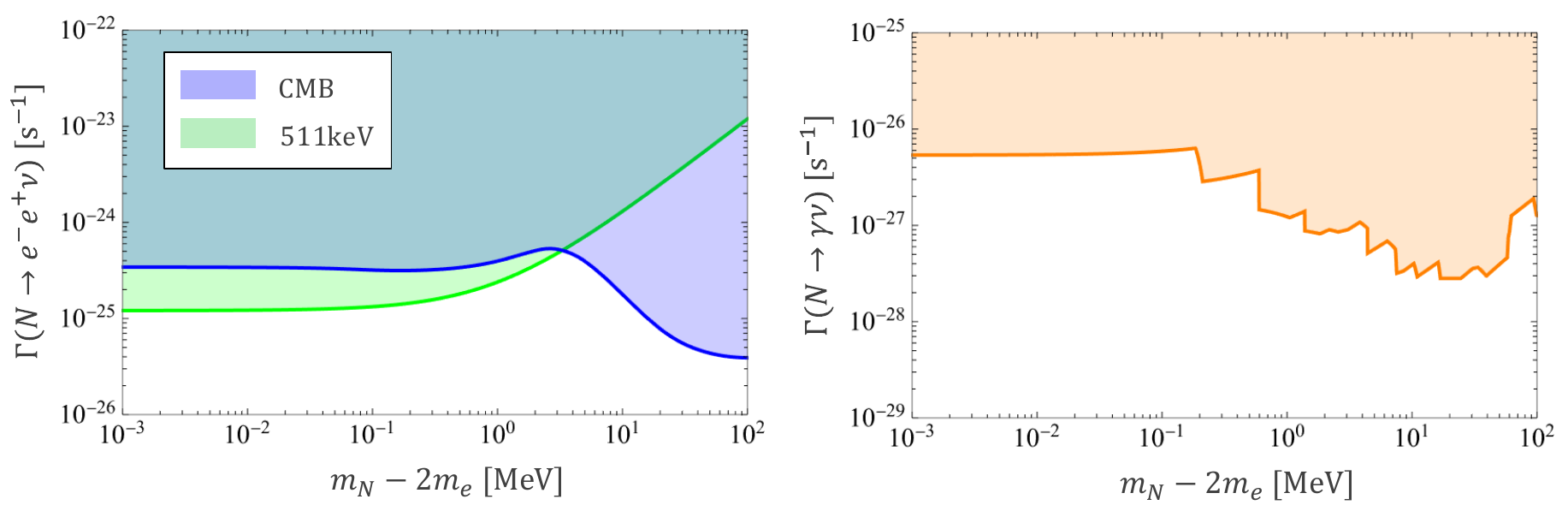}
    \caption{\small \sl \textbf{Left panel:} Upper limits on the width $\Gamma[N \to e^- e^+ \nu]$ are shown as a function of $m_N - 2m_e$, derived from the CMB observations and the 511\,keV gamma-ray line observed from the Galactic center. \textbf{Right panel:} Upper limit on $\Gamma(N \to \gamma \nu)$ derived from gamma-ray observations is shown.}
    \label{fig: constraints}
\end{figure}

The decay process $N \to e^- e^+ \nu$ affects the thermal history of the universe, particularly during the epochs of recombination\,\cite{Slatyer_2017} and reionization\,\cite{Liu_2016}. However, cosmological observations favor a thermal history without such energy injections, thereby placing constraints on the decay width of sterile neutrino dark matter decaying into electrons. Among various constraints, the most stringent is expected to arise from precise observations of the CMB. Although no direct limit on the process $N \to e^- e^+ \nu$ based on the latest data has been reported in the literature, an indirect constraint can be inferred from the CMB limit on decays into a monochromatic electron-positron pair, $V \to e^- e^+$. The corresponding constraint on the decay width, $\Gamma[V \to e^- e^+](m_V)$ (i.e., the lifetime of $V$), is presented as a function of the particle mass $m_V$ in Fig.\,7 of Ref.\,\cite{Slatyer_2017}. In contrast, sterile neutrino dark matter does not decay into a monochromatic electron-positron pair, but instead produces an electron and a positron with a continuous energy spectrum, $d\Gamma[N \to e^- e^+ \nu]/dE_{e^\pm}$, whose explicit form is provided in Appendix\,\ref{app: Sommerfeld effect}. Taking into account that each decay produces both an electron and a positron sharing the same spectrum, the constraint on the decay width is obtained as
\begin{align}
    \Gamma[N \to e^- e^+ \nu] \leq
    \int_{m_e}^{\frac{m_N}{2}} dE_{e^\pm}\,
    \frac{1}{\Gamma[N \rightarrow e^- e^+ \nu]} \frac{d\Gamma[N \rightarrow e^- e^+ \nu]}{dE_{e^\pm}}
    \Gamma[V \to e^-e^+](2E_{e^\pm}).
    \label{eq: differential decay width}
\end{align}
The resulting constraint on the partial decay width $\Gamma[N \to e^- e^+ \nu]$ at the 95\,\% confidence level is presented in the left panel of Fig.\,\ref{fig: constraints} as a function of $m_N - 2m_e$. Assuming $\Theta_{11} = \Theta_{21} = \Theta_{31} = \Theta$, this constraint can be translated into a bound on $\Theta^2$, as shown in Fig.\,\ref{fig: theta}.

%%%%%%%%%%%%%%%%%%%%%%%%%%%%%%%%%
%%%%%%%%%%% Detection %%%%%%%%%%%
%%%%%%%%%%%%%%%%%%%%%%%%%%%%%%%%%
\section{Detection of the sterile neutrino dark matter}
\label{sec: detection}

The sterile neutrino dark matter is expected to be detected through astrophysical observations, i.e., indirect detection. Its decay produces various observable signals, such as neutrinos, electrons, positrons, and photons. Although neutrinos constitute the majority of the decay products, their detection is severely limited by the intrinsically low efficiency of neutrino detectors. Given the cosmological constraints on $\Theta_{i1}$ discussed in the previous section, the predicted signal strength remains well below the sensitivity of current observations\,\cite{Alvi_2022}. In addition, the detection of electrons and positrons is inefficient for MeV-scale $m_N$, as their low energies prevent them from penetrating the heliosphere. While Voyager~1, located outside the heliosphere, could in principle detect such low-energy particles, its sensitivity remains insufficient for electrons and positrons with MeV-scale energies\,\cite{Boudaud_2017}.

Detecting photons, i.e., gamma-ray signals in the context of indirect detection, is easier than detecting neutrinos, electrons, or positrons. Sterile neutrino dark matter can produce observable signals via the following processes. The first is final-state radiation emitted during the decay into a pair of electrons and a neutrino, i.e., $N \to e^- e^+ \nu\,\gamma$. The second is radiative decay into a photon and a neutrino, i.e., $N \to \gamma \nu$. The third, known as \textit{secondary production}, arises from interactions between the electrons and positrons produced in the decay and the interstellar medium, including inverse Compton scattering off background radiation and Bremsstrahlung in interstellar gas. The signal strength from the third process is suppressed for the MeV-scale $m_N$, since the injected electrons and positrons generate secondary photons with insufficiently high energy.\footnote{
    The secondary contribution depends sensitively on how low-energy electrons and positrons propagate in our galaxy. Although their propagation is not yet well understood, the signal could be enhanced if reacceleration is efficient\,\cite{luque2024importance}. We do not consider such a boost here in order to keep the constraints conservative.}
Consequently, the resulting signal is not strong enough to exceed the cosmological bounds discussed in the previous section\,\cite{Cirelli:2020bpc, Cirelli:2023tnx}.

\begin{figure}[t]
    \centering
    \includegraphics[keepaspectratio, scale=0.6]{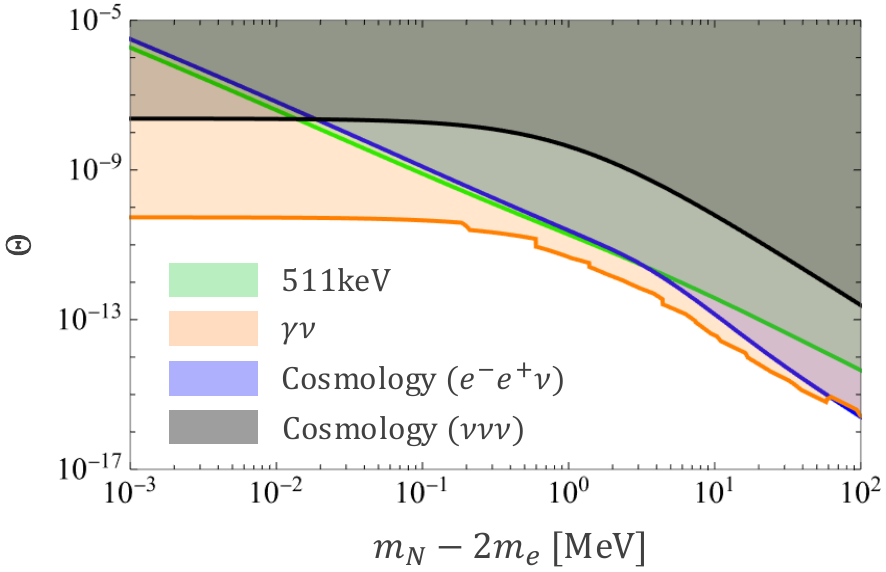}
    \caption{\small \sl Upper limits on the mixing angle between the lightest right-handed neutrino and the left-handed neutrinos, $\Theta_{i1} = y_{i1} v_{\rm EW}/(\sqrt{2}\,m_N)$, are shown as a function of $m_N - 2m_e$. These limits are derived from cosmological observations ($N \to \nu\nu\nu$, $e^-e^+\nu$), the 511\,keV gamma-ray line, and the gamma-ray line at $E_\gamma = m_N/2$ ($N \to \gamma \nu$), under the assumption $\Theta_{11} = \Theta_{21} = \Theta_{31} \equiv \Theta$.}
    \label{fig: theta}
\end{figure}

The signal from the first process (i.e., the internal bremsstrahlung process) is weaker than that from the second process (i.e., the radiative decay), as the former is suppressed by the four-body final state, while the latter is suppressed by a one-loop factor. Moreover, the radiative decay produces a distinctive signature, a monochromatic photon line, against astrophysical backgrounds, thereby enabling a more effective search for the sterile neutrino dark matter. The partial decay width for the radiative process $N \to \gamma \nu$ is given by
\begin{align}
    &
    \Gamma(N \to \gamma \nu) = \frac{\alpha\,G_F^2}{64\,\pi^4} m_N^5 \sum_a \Theta_{a1}^2\,
    F^2\left(
        \frac{m_{l_a}^2}{m_W^2},\,\frac{m_N^2}{m_W^2}
    \right) \simeq
    \frac{9 \alpha\,G_F^2}{256\,\pi^4} m_N^5 \sum_a \Theta_{a1}^2,
    \\
    &
    F(x,y) =
    \int_0^1 ds \int_0^{1-s} dt \left[
        \frac{2s(s+t) + (1 - 2s + s^2 + st)x}
        {s + (1-s)x - s(1-s-t)y}
        +\frac{2(1 - 2s + s^2 + st)+ s(s+t)x}
        {1-s + sx - s(1-s-t)y}
    \right],
    \nonumber
\end{align}
where $\alpha$ is the fine-structure constant and $m_{l_a}$ is the mass of the charged lepton $l_a$. We use the approximation $m_{l_a}, m_N \ll m_W$ to derive the final expression for the decay width. Since such a monochromatic signal has not been observed by current X-ray and $\gamma$-ray observatories\,\cite{Essig:2013, Laha:2020}, its absence places an upper bound on the partial decay width $\Gamma(N \to \gamma \nu)$, as shown in the right panel of Fig.\,\ref{fig: constraints} as a function of $m_N - 2m_e$. This constraint can be translated into an upper bound on $\Theta^2$, as shown in Fig.\,\ref{fig: theta}, under the assumption that $\Theta_{11} = \Theta_{21} = \Theta_{31} \equiv \Theta$.

In addition to the signals mentioned above, the decay of the sterile neutrino dark matter produces another distinctive photon signal, known as \textit{tertiary production}. Most of the positrons produced in the decay, which have MeV-scale energies, form positronium by capturing ambient electrons. The positronium subsequently decays into two photons with a branching fraction of approximately 25\%, resulting in the characteristic 511\,keV photon signal. Unlike the other signals, the 511\,keV photon line has already been observed; its flux from the Galactic bulge region is measured to be $F_{511}^{\rm Obs} \simeq (4.8$--$9.6) \times 10^{-4}$\,ph\,cm$^{-2}$\,s$^{-1}$\,\cite{Siegert_2016}.\footnote{
    Here, the bulge region is defined as a circular area centered on the Galactic center with a radius of $10.3^\circ$.}
On the other hand, the origin of the observed flux is not well understood; it may arise from certain astrophysical sources, such as microquasars, supernovae, or massive stars\,\cite{dissertation, Siegert_2023}, rather than from dark matter decay. Therefore, the observed 511\,keV flux places the upper bound on the partial decay width into an electron-positron pair, as given in Eq.\,(\ref{eq: eenu}):
\begin{align}
    F_{511}^{\rm DM} =
    2 \times
    \frac{1}{4} \times
    f_{Ps} \times
    {\cal E}_{ff}
    \times
    \frac{\Gamma(N \to e^-e^+\nu)}{4\pi m_N}\,D
    < F_{511}^{\rm Obs},
    \label{eq: const from 511}
\end{align}
where the factors of 2 and $1/4$ arise from the number of photons emitted per positronium decay and the branching fraction of positronium decaying into two photons. The parameter $f_{Ps}$ represents the fraction of $e^- e^+$ annihilations that contribute to the 511\,keV flux via positronium formation, which is found to be consistent with 100\,\% within observational uncertainties\,\cite{Siegert_2016}. The $D$-factor is obtained by integrating the dark matter mass density profile $\rho_{\rm DM}(\vec{r})$ along the line of sight (l.o.s.) in the direction of the bulge region $\Omega$, and is defined as $D \equiv \int_{\rm Bulge} d\Omega \int_{\rm l.o.s.} d\ell\, \rho_{\rm DM}(\ell, \Omega)$. Here, we assume a spherically symmetric dark matter profile, $\rho_{\rm DM}(\vec{r}) = \rho_{\rm DM}(r)$, with $r$ denoting the distance from the Galactic center. We adopt a conservative value of the $D$-factor, $3.4 \times 10^{21}\,\mathrm{GeV/cm^2}$, following Ref.\,\cite{Hayashi:2024}. We also include the efficiency factor ${\cal E}_{\rm ff}$, which accounts for the possibility that a fraction of the positrons produced within the bulge region may escape without undergoing annihilation. This factor is subject to significant uncertainty, with estimates ranging from 6\,\% to 100\,\%\,\cite{dissertation, Siegert_2023}; to remain conservative, we adopt the lowest value in our analysis. An upper bound on the partial decay width $\Gamma(N \to e^-e^+\nu)$ is then obtained from the constraint\,(\ref{eq: const from 511}) and is shown in the left panel of Fig.\,\ref{fig: constraints} as a function of $m_N - 2m_e$, while the corresponding constraint on the angle $\Theta^2$ is presented in Fig.\,\ref{fig: theta}, under the assumption that $\Theta_{11} = \Theta_{21} = \Theta_{31} \equiv \Theta$.\footnote{
    Refs.\,\cite{DelaTorreLuque:2023cef, Nguyen:2025tkl} shows that morphology data strengthen the constraint. We instead adopt a more conservative bound, avoiding the morphology due to its non-negligible uncertainties, as also noted in the references.}

We are in a position to discuss the future prospects for probing sterile neutrino dark matter through upcoming MeV gamma-ray observations. Our focus is particularly on the \textit{COSI} mission, which is scheduled for launch in 2027\,\cite{Tomsick:2021wed} as a NASA SMEX mission. A key advantage of COSI is its excellent energy resolution in the 200\,keV to 5\,MeV range, enabling the potential detection of sharp monochromatic gamma-ray lines, such as the 511\,keV line and the line at $m_N/2$, as discussed above. As usual, we consider the Galactic Center region as the region of interest (ROI) for detecting the line signal at $m_N/2$\,\cite{Essig:2013}. In contrast, for the 511\,keV line originating from dark matter decay, we adopt the entire region outside the Galactic bulge as the ROI. This choice is motivated by the following consideration. In decaying dark matter scenarios, the observed 511\,keV flux cannot be attributed to dark matter, as its morphology differs significantly from that expected for a dark matter origin. The observed emission is therefore likely dominated by astrophysical sources such as microquasars, supernovae, and massive stars. Since the astrophysical background is concentrated in the bulge region, whereas the dark matter signal is generally more spatially extended, focusing on the region outside the bulge offers improved sensitivity to a potential signal. It is also worth noting that this strategy becomes viable only for missions employing a Compton camera, such as COSI. The expected signal would contribute as an almost isotropic diffuse component of the 511\,keV flux, making its detection feasible with a Compton camera, but generally difficult with previous telescopes using a coded-aperture mask\,\cite{Tomsick:2021wed, Aramaki:2022zpw, COSI2023}.

\begin{figure}[t]
    \centering
    \includegraphics[keepaspectratio, scale=0.35]{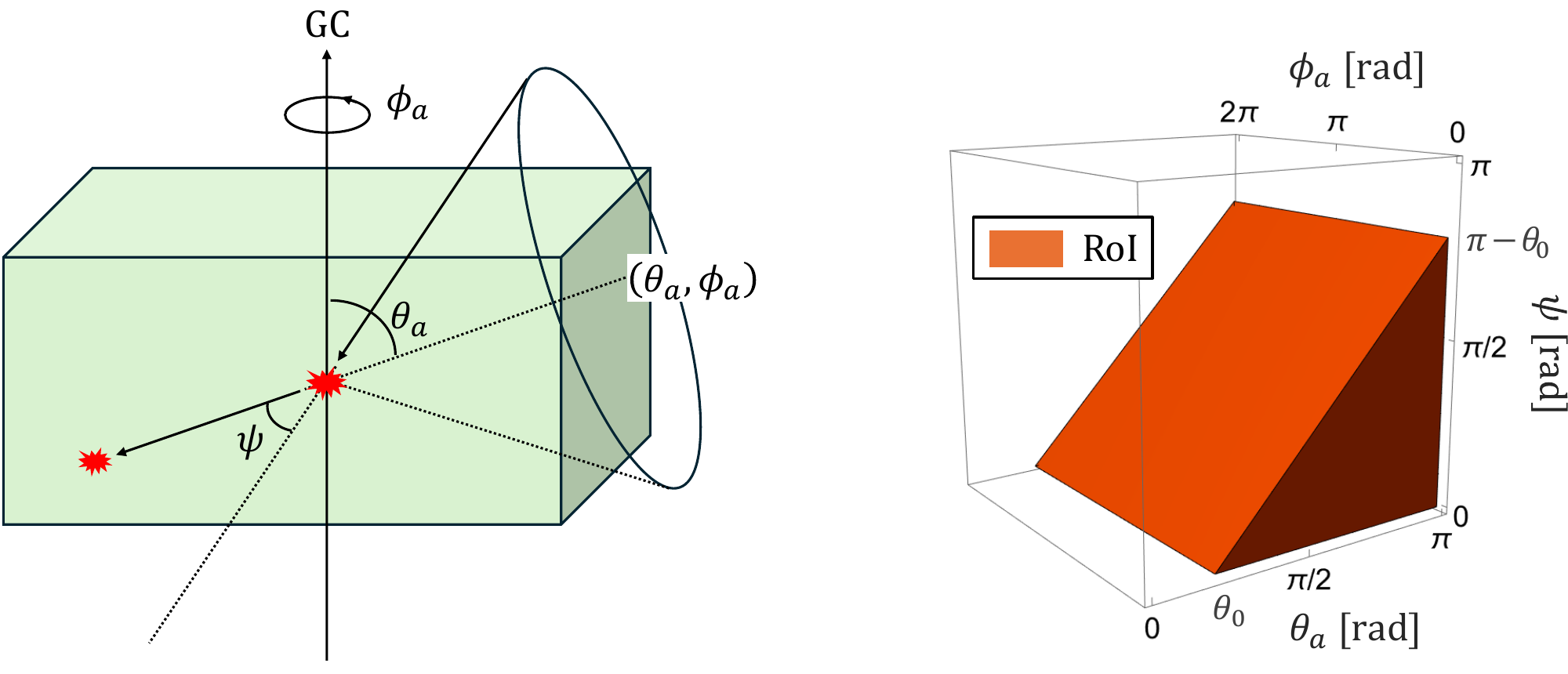}
    \caption{\small \sl {\bf Left:} Definition of the Compton scattering angle $\psi$ and the apparent direction $(\theta_a, \phi_a)$ of a 511\,keV event, as observed in the detector. {\bf Right:} The Compton data space spanned by the scattering angle and the apparent direction. Each observed event is recorded as a point in this space. The red region shown in the figure represents the region of interest (RoI) adopted in our analysis.}
    \label{fig: detector}
\end{figure}

\begin{figure}[t]
    \centering
    \includegraphics[keepaspectratio, scale=0.6]{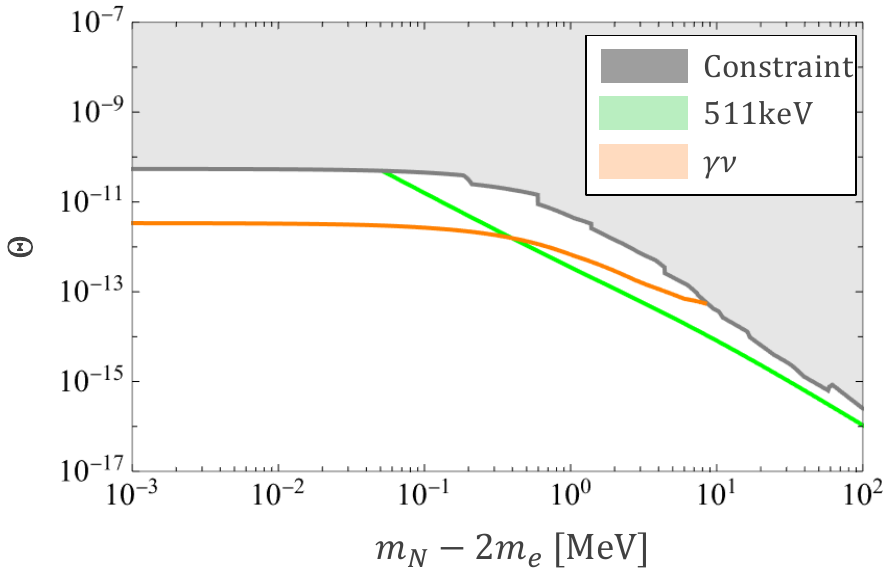}
    \caption{\small\sl The sensitivity of the COSI observation to the mixing angle between the lightest right-handed neutrino and the left-handed neutrinos, defined as $\Theta_{i1} = y_{i1} v_{\rm EW}/(\sqrt{2}\,m_N)$, is shown as a function of $m_N - 2m_e$, under the assumption $\Theta_{11} = \Theta_{21} = \Theta_{31} \equiv \Theta$. The sensitivity is derived from searches for the $N \to \gamma \nu$ line signal from the Galactic center region and the 511\,keV signal originating from the $N \to e^- e^+ \nu$ decay in regions outside the Galactic bulge. The parameter space already excluded by current observations (shown in Fig.\,\ref{fig: theta}) is indicated by the gray shaded region.}
    \label{fig: future}
\end{figure}

To quantitatively estimate the sensitivity of the COSI observation to a monochromatic gamma-ray line signal at $E_\gamma = m_N/2$, we follow the procedure described in Ref.\,\cite{Caputo:2022}. Specifically, we utilize the publicly available COSI sensitivity data\,\cite{COSI2023}, provided as a function of photon energy, to evaluate the expected sensitivity to a line signal originating from the Galactic center region with a radius of $10^\circ$. In this analysis, we assume a total observation time of 24 months. The resulting sensitivity is shown as the orange solid line in Fig.\,\ref{fig: future}.

To evaluate the COSI sensitivity to the 511\,keV line signal arising from dark matter decay, we define the region of interest (ROI) as the entire sky excluding the bulge region, which is modeled as a circular area with a $\psi$-dependent angular radius $\theta_{\rm ROI}(\psi)$. Here, $\psi$ denotes the Compton scattering angle of the incoming photon, as measured by the COSI detector (Fig.\,\ref{fig: detector}). This $\psi$-dependent definition of the ROI accounts for the complexities of event reconstruction in the Compton camera and is motivated by the following considerations:
\begin{itemize}
    \item
    The observed 511\,keV flux, which is expected to be predominantly due to astrophysical backgrounds, appears to follow the density profile $n_{511}(r) = n_{\rm BB}(r) + n_{\rm NB}(r) + n_{\rm c}(r)$:
    \begin{align}
        n_{\rm BB}(r) &= f_{\rm BB}\,e^{-r^2/(2\sigma_{\rm BB}^2)}, \quad
        n_{\rm NB}(r) = f_{\rm NB}\,e^{-r^2/(2\sigma_{\rm NB}^2)}, \quad
        n_{\rm c}(\vec{r}) = f_{\rm c}\,\delta(\vec{r}),
    \end{align}
    where the functions $n_{\rm BB}(r)$, $n_{\rm NB}(r)$, and $n_{\rm c}(\vec{r})$ denote the broad, narrow, and central bulge components, respectively, as inferred from the 511\,keV observations\,\cite{Skinner:2015}. The corresponding parameters are $(f_{\rm BB}, \sigma_{\rm BB}) = (3.80 \times 10^{-2}\,\mathrm{cm^{-2} \, s^{-1} \, kpc^{-1}}, 1.26\,\mathrm{kpc})$, $(f_{\rm NB}, \sigma_{\rm NB}) = (3.08 \times 10^{-1}\,\mathrm{cm^{-2} \, s^{-1} \, kpc^{-1}}, 0.363\,\mathrm{kpc})$, and $f_{\rm c} = 8.0 \times 10^{-5}\,\mathrm{cm^{-2} \, s^{-1}}$. Using this density profile, the observed 511\,keV flux is then expressed as follows:
    \begin{align}
        \frac{d{\cal F}_{511}^{\rm Obs}(\theta)}{d\Omega} =
        \frac{1}{4\pi} \int_{\rm l.o.s} d\ell\,n_{511}(r),
    \end{align}
    where $r^2 = \ell^2 - 2\ell d_{\rm GC} \cos \theta + d_{\rm GC}^2$, $\Omega = \Omega(\theta, \phi)$ defines the direction of the line of sight, and $d_{\rm GC}$ is the distance between the Galactic center and the Solar System.

    \item
    We assume that the morphology of the signal flux is proportional to the dark matter density profile $\rho_{\rm DM}(r)$, and adopt a spherically symmetric, cored profile given by
    \begin{align}
        \rho_{\rm DM}(r) = \frac{\rho_s}{(1+r/r_s)[1+(r/r_s)^2]} \,,
    \end{align}
    where $\rho_s \simeq 0.71$\,GeV/cm$^3$ and $r_s = 12.7$\,kpc. This is the same profile adopted in Eq.\,(\ref{eq: const from 511}) to evaluate the constraint on the width $\Gamma[N \to e^- e^+ \nu]$ from the observed 511\,keV flux.  Then, the 511\,keV flux from the dark matter decay is obtained as
    \begin{align}
        \frac{d {\cal F}_{511}^{\rm DM}(\theta)}{d\Omega} \simeq
        2 \times \frac{1}{4} \times
        \frac{\Gamma(N \to e^- e^+ \nu)}{4 \pi m_N}
        \int_{\rm l.o.s} d\ell\,\rho_{\rm DM}(r),
    \end{align}
     The assumption about the morphology of the signal flux implies that we neglect the effect of positron propagation after their production via dark matter decay. Since we define the ROI as the region outside the bulge to search for the 511\,keV signal, including the propagation effect could potentially enhance the signal, as positrons produced within the bulge may propagate outward and form positronium, which subsequently decays into 511\,keV photons. However, the propagation of positrons in the bulge, especially those with low energies, is still under debate\,\cite{Siegert_2023}. Therefore, we do not incorporate this effect in the present analysis and leave it for future investigation.
\end{itemize}

Then, we define the angular radius $\theta_0$ of the bulge region in the coordinates shown in Fig.\,\ref{fig: detector} as the angle at which the dark matter signal becomes comparable to the astrophysical background, that is, when ${\cal F}_{511}(\theta_0, \phi) - {\cal F}^{\rm DM}_{511}(\theta_0, \phi) = {\cal F}^{\rm DM}_{511}(\theta_0, \phi)$. Next, we introduce the $\psi$-dependent angular radius $\theta_{\rm ROI}(\psi)$, taking into account the characteristics of Compton cameras, which are employed in several next-generation MeV gamma-ray observatories such as COSI. In observations using a Compton camera, events are recorded as points in the Compton data space, defined by the observed apparent coordinates $(\theta_a, \phi_a)$ and the Compton scattering angle $\psi$, as illustrated in Fig.\,\ref{fig: detector}. As shown in the right panel, the angular extent of the bulge region defined by $\theta_0$ at $\psi = 0$ effectively increases with larger values of $\psi$, such that $\theta_{\rm ROI}(\psi) = \theta_0 + \psi$. Accordingly, the signal flux within the ROI is given by
\begin{align}
    &{\cal F}^{\rm DM}_{511} =
    (2\pi)
    \int_0^{\pi - \theta_0} d\psi\,P(\psi)
    \int_{\theta_{\rm ROI}(\psi)}^\pi \sin\theta\,d\theta\,\frac{{\cal F}_{\rm DM}^{511}(\theta)}{d\Omega},
    \nonumber \\
    &P(\psi) =
    \frac{9 \sin\psi}{40 - 27\log 3}
    \frac{2 + (1-\cos\psi)^3}{(2-\cos\psi)^3}.
\end{align}
Here, the probability function $P(\psi) = \sigma_{\rm KN}^{-1}\,(d\sigma_{\rm KN}/d\psi)$, where $\sigma_{\rm KN}$ is the Klein–Nishina cross-section, evaluated for an incoming photon energy set by the electron mass, i.e., $E_\gamma = 511\,\mathrm{keV}$. According to the publicly available COSI sensitivity data for gamma-ray line observations\,\cite{COSI2023}, the minimum detectable 511\,keV flux is ${\cal F}^{\rm limit}_{511} = 1.2 \times 10^{-5}\,\mathrm{cm}^{-2}\,\mathrm{s}^{-1}$ at the $3\sigma$ confidence level, assuming 24 months of observation. For a rough estimate of the COSI sensitivity to 511\,keV signals from dark matter decay outside the bulge region (i.e., in the RoI region), we also plot the line ${\cal F}_{511}^{\rm DM} = {\cal F}^{\rm limit}_{511}$ in Fig.\,\ref{fig: future}. To evaluate the sensitivity in a more rigorous manner, in addition to the effect of positron propagation mentioned above, the RoI adopted in this article should be replaced by a more sophisticated one that avoids background contributions to the 511\,keV signal count not only from the bulge region but also from the Galactic disk and bright sources in the Compton data space. Alternatively, it may be possible to retain the same RoI as used in this article, but include the background contributions from the Galactic disk and bright sources in the likelihood analysis. The choice of analysis procedure, i.e., how to efficiently extract the 511\,keV signal from dark matter decay, requires a detailed response model of the Compton camera, which is still under development, and is beyond the scope of this study. Therefore, we leave it for future work.

As illustrated in Fig.\,\ref{fig: future}, the monochromatic gamma-ray line signal at $E_\gamma = m_N/2$, produced via the radiative decay process $N \to \gamma \nu$ in the Galactic center region, can be effectively probed by observations with the COSI instrument. This detection channel presents a promising opportunity to explore a previously inaccessible region of the sterile neutrino parameter space, as already pointed out in the literature\,\cite{Caputo:2022}. The sensitivity of this search is particularly enhanced in the low-mass regime, specifically for $m_N \lesssim 10$\,MeV (corresponding to $E_\gamma \lesssim 5$\,MeV), due to COSI's excellent energy resolution in this range. However, for larger values of $m_N$, the sensitivity decreases, and the capacity to constrain new parameter space becomes increasingly limited. On the other hand, the 511\,keV gamma-ray signal arising from the three-body decay process $N \to e^- e^+ \nu$, when observed from regions outside the Galactic bulge, provides a complementary probe. This channel enables the exploration of a distinct and previously untested region of parameter space, extending beyond $m_N \lesssim 10$\,MeV up to ${\cal O}(100)$\,MeV within the reach of COSI observations. The extended sensitivity in this mass range results from the distinct spatial morphology of the signal outside the bulge, which reduces contamination from known astrophysical backgrounds. Moreover, it is noteworthy that both decay channels, $N \to \gamma \nu$ and $N \to e^- e^+ \nu$, can, in principle, be simultaneously probed in a single observation, provided the sterile neutrino mass satisfies $50\,\mathrm{keV} \leq m_N - 2m_e \leq 10\,\mathrm{MeV}$. Within this overlapping mass window, the simultaneous observation of a sharp gamma-ray line and the 511\,keV emission would constitute a striking and correlated signature of sterile neutrino dark matter decay, thereby offering a unique opportunity to test this scenario with upcoming MeV gamma-ray missions.

%%%%%%%%%%%%%%%%%%%%%%%%%%%%%%%%%%%%%%%%%%%%
%%%%%%%%%%% Summary & Discussion %%%%%%%%%%%
%%%%%%%%%%%%%%%%%%%%%%%%%%%%%%%%%%%%%%%%%%%%
\section{Summary \& Discussion}
\label{sec: summary and future direction}

The existence of non-zero neutrino masses, dark matter, and the baryon-number asymmetry in the Universe points to the need for physics beyond the SM. A minimal and theoretically appealing extension is the inclusion of three right-handed neutrinos. In particular, the gauged \( U(1)_{\rm B-L} \) model naturally accommodates this extension: two of the right-handed neutrinos can explain neutrino masses via the seesaw mechanism and generate the baryon asymmetry through leptogenesis, while the third serves as a dark matter candidate in the form of a sterile neutrino. This work investigates the indirect detection of sterile neutrino dark matter through its decay channels. The radiative decay \( N \to \gamma \nu \) produces a monochromatic photon with energy \( E_\gamma = m_N/2 \), while the three-body decay \( N \to e^- e^+ \nu \) results in 511\,keV photons via positron annihilation. We focus on the observational prospects for these signals using future MeV gamma-ray missions, taking COSI as a representative example.

We demonstrate that COSI possesses excellent sensitivity to the $N \to \gamma \nu$ line signal for dark matter masses below 10\,MeV, owing to its superior energy resolution. For larger masses, the sensitivity to this channel diminishes; however, the 511\,keV signal from $N \to e^- e^+ \nu$, particularly when observed from regions outside the Galactic bulge, provides a complementary probe extending up to ${\cal O}(100)\,\mathrm{MeV}$. This extended reach is enabled by the more isotropic morphology of the signal and the reduced contamination from known astrophysical backgrounds in such regions. To optimally extract the 511\,keV signal, we propose a novel analysis strategy based on defining the region of interest in Compton data space. In the overlapping mass range $50\,\mathrm{keV} \leq m_N - 2m_e \leq 10\,\mathrm{MeV}$, both the monochromatic line and 511\,keV signals can be observed simultaneously. Since both are governed by the same neutrino Yukawa coupling, their correlated detection would provide a distinctive and testable signature of sterile neutrino dark matter. Moreover, we compute the decay width of \( N \to e^- e^+ \nu \) including, for the first time, the Sommerfeld enhancement arising from threshold singularities.

To further refine the analysis presented in this work, it will be essential to incorporate the propagation effects of MeV-scale positrons in the Galaxy, particularly in the bulge region, as well as to improve the treatment of astrophysical backgrounds in the Compton data space. The former requires precise modeling of the interstellar medium and radiation fields within the Galaxy, especially in the bulge, while the latter calls for an accurate response model of the MeV gamma-ray observatories based on Compton camera technology. Both aspects are currently under active investigation, and incorporating them into future analyses will be crucial for achieving more robust and reliable projections of sterile neutrino dark matter signals in upcoming MeV gamma-ray missions. On the other hand, when focusing on regions outside the Galactic bulge, one must also consider potential contributions to the 511\,keV gamma-ray signal not only from dark matter decay in the Galactic halo, as emphasized in this study, but also from extragalactic dark matter or astrophysical sources. Therefore, if a diffuse, isotropic 511\,keV signal is detected by next-generation MeV gamma-ray observations (a remarkable discovery in its own right), establishing its true origin will be of critical importance. In this context, a comparative analysis between the 511\,keV signal and the monochromatic gamma-ray line from the decay process $N \to \gamma \nu$ becomes essential. In particular, the relative intensities of the line signal from the Galactic center, the bulge, and the outer regions, when compared with the corresponding 511\,keV emission, provide valuable insights into the spatial distribution and origin of the underlying dark matter population.

%%%%%%%%%%%%%%%%%%%%%%%%%%%%%%%%%%%%%%%
%%%%%%%%%%% Acknowldgements %%%%%%%%%%%
%%%%%%%%%%%%%%%%%%%%%%%%%%%%%%%%%%%%%%%
\section*{Acknowledgments}

All authors were supported by the World Premier International Research Center Initiative (WPI), MEXT, Japan (Kavli IPMU). S.~M. was supported by the Grant-in-Aid for Scientific Research from the Ministry of Education, Culture, Sports, Science and Technology, Japan (MEXT), under Grant Nos. 24H00244 and 24H02244.

%%%%%%%%%%%%%%%%%%%%%%%%%%%%%%%%%%
%%%%%%%%%%% Appendices %%%%%%%%%%%
%%%%%%%%%%%%%%%%%%%%%%%%%%%%%%%%%%
\appendix

\section{Sommerfeld effect}
\label{app: Sommerfeld effect}

To calculate the decay width including the effect, we consider the decomposition of the width in terms of the partial waves of the $e^-e^+$ system. In the threshold region, the decay into $e^-e^+\nu$ can be regarded as a decay into an $e^-e^+$ pair and a neutrino. The non-relativistic $e^-e^+$ system can be decomposed as $(e^-e^+) = (e^-e^+)_{{}^1S_0} + (e^-e^+)_{{}^3S_1} + (e^-e^+)_{{}^1P_1} + (e^-e^+)_{{}^3P_0} + (e^-e^+)_{{}^3P_1} + \cdots$, where the Russell–Saunders term symbols ${}^{2S+1}L_J$ (with $S$, $L$, and $J$ denoting the spin, orbital, and total angular momenta, respectively) are used to specify the quantum numbers of the two-body system. As shown below, the dominant contributions to the decay width come from the decay modes into $(e^-e^+)_{{}^1S_0}$ and $(e^-e^+)_{{}^3S_1}$. In the perturbative calculation (i.e., without the Sommerfeld effect), their partial decay widths are given by
\begin{align}
    &\Gamma_{\rm NR}[N \to (e^-e^+)_{{}^1S_0} \nu] \simeq
    \frac{G_F^2 m_N^5}{32\pi^3} \left[
        \sum_a \Theta_{a1}^2
    \right]\,I^{(0)}\left(\frac{m_N}{m_e}\right),
    \label{eq: NR eenu 0}
    \\
    &\Gamma_{\rm NR}[N \to (e^-e^+)_{{}^3S_1} \nu] \simeq \frac{3 G_F^2 m_N^5}{32\pi^3} \left[
        \sum_a \Theta_{a1}^2 (4s_W^2 + 2\delta_{a1} -1)^2
    \right]\,I^{(0)}\left(\frac{m_N}{m_e}\right),
    \label{eq: NR eenu 1}
\end{align}
where we take the non-relativistic limit for the electron-positron pair, i.e., $m_N \simeq 2 m_e$, to derive the above expressions for the decay widths. Here, the function $I(x)$ is given by
\begin{align}
    I^{(0)}(x) =
    \int_0^{x-2} dy\, \frac{y^{1/2} (\sqrt{x - 1 - y} - 1)^2}{2 \sqrt{x - 1 - y}}.
    \label{eq: I0(x)}
\end{align}
The comparison between the non-relativistic decay width, $\Gamma_{\rm NR}(N \to e^-e^+\nu)$, defined as  
\begin{align}
    \Gamma_{\rm NR}(N \to e^-e^+\nu) \equiv
    \Gamma_{\rm NR}[N \to (e^-e^+)_{{}^1S_0} \nu] + \Gamma_{\rm NR}[N \to (e^-e^+)_{{}^3S_1} \nu],
\end{align}  
and the full tree-level decay width, $\Gamma(N \to e^-e^+\nu)$, given in Eq.\,(\ref{eq: eenu}), demonstrates that this non-relativistic contribution, $\Gamma_{\rm NR}(N \to e^-e^+\nu)$, dominates near the threshold, as expected. The difference between the two widths originates from higher-order non-relativistic corrections to the decay widths into $(e^-e^+)_{{}^1S_0}\nu$ and $(e^-e^+)_{{}^3S_1}\nu$, as well as from additional decay channels, such as $N \to (e^-e^+)_{{}^1P_1}\nu$, which are not captured in the leading-order non-relativistic approximation. The contributions from these higher partial waves are suppressed in the threshold region due to the small relative velocity of the $e^-e^+$ pair, regardless of whether the Sommerfeld enhancement is included or not. In contrast, the channels $N \to (e^-e^+)_{{}^1S_0}\nu$ and $N \to (e^-e^+)_{{}^3S_1}\nu$ provide the dominant contributions in the non-relativistic regime. We therefore focus on the Sommerfeld effect in these two channels.

The Sommerfeld effect in these channels can be evaluated using the so-called potential non-relativistic (pNR) Lagrangian\,\cite{Pineda_1998, Brambilla_2000}. This is an effective field theory derived by integrating out the hard ($\ell^0 \sim |\vec{\ell}| \sim m_e$) and soft ($\ell^0 \sim |\vec{\ell}| \sim \beta m_e$) modes, while retaining the potential ($\ell^0 \sim \beta^2 m_e$, $|\vec{\ell}| \sim \beta m_e$) and ultrasoft ($\ell^0 \sim |\vec{\ell}| \sim \beta^2 m_e$) components of the electron and photon fields from the original Lagrangian. Here, $\beta$ denotes the velocity of the electron. Starting from the Lagrangian\,(\ref{eq: interactions 2 II}), one integrates out all fields except those associated with sterile neutrino dark matter, active neutrinos, and the potential-mode electron field $e_{\rm pot}(x)$.\footnote{
    Since our focus is on the leading NR contribution to $\Gamma(N \to e^- e^+ \nu)$, we further integrate out the ultra-soft component of the electron field $e(x)$ as well as all components of the photon field from the Lagrangian.} The field $e_{\rm pot}(x)$ is then expanded systematically in powers of its velocity as
\begin{align}
    e_{\rm pot}(x) =
    \begin{pmatrix}
        e^{- i m_e x^0} \eta(x) + i e^{i m_e x^0} [\vec{\nabla} \cdot \vec{\sigma}\,\xi(x)]/(2 m_e) + \cdots \\
        e^{i m_e x^0} \xi(x) - i e^{-i m_e x^0} [\vec{\nabla} \cdot \vec{\sigma}\,\eta(x)]/(2 m_e) + \cdots
    \end{pmatrix},
    \label{eq: NR expansion}
\end{align}
with $\vec{\sigma}$ being the Pauli matrices, giving us the non-relativistic Lagrangian as follows:
\begin{align}
    \mathcal{L}_{\rm NR}
    &\simeq
    \frac{1}{2} \bar{N} (i \slashed{\partial} + m_N) N
    +
    \sum_a \bar{\nu}_a i \slashed{\partial} \nu_a
     +
    \eta^\dagger \left( i\partial_{x^0} + \frac{\nabla^2}{2m_e} \right) \eta
    +
    \xi^\dagger \left( i\partial_{x^0} - \frac{\nabla^2}{2m_e} \right) \xi
    \nonumber \\
    &+
    \int d^4y \frac{\alpha\,\delta(x^0 - y^0)}{2\,|\vec{x}-\vec{y}|}
    \left(
        \left[\eta^\dagger(x) \xi(y)\right]
        \left[\xi^\dagger(y) \eta(x)\right]
        +
        \left[\eta^\dagger(x)\,\vec{\sigma}\,\xi(y)\right]
        \cdot
        \left[\xi^\dagger(y)\,\vec{\sigma}\,\eta(x) \right]
    \right)
    \nonumber \\
    &+
    \sum_a \frac{G_F\,\Theta_{a1}}{\sqrt{2}}
    (2\delta_{a1} -1) (\bar{\nu}_a \gamma^0 N)
    (e^{2im_e x^0} \eta^\dagger \,\xi + e^{-2im_e x^0} \xi^\dagger \,\eta) + h.c.
    \nonumber \\
    &+
    \sum_a \frac{G_F\,\Theta_{a1}}{\sqrt{2}}
    (4s_W^2 + 2\delta_{a1} -1)
    (\bar{\nu}_a \vec{\gamma} N) 
    \cdot
    (e^{2im_e x^0} \eta^\dagger\vec{\sigma}\xi + e^{-2im_e x^0} \xi^\dagger\vec{\sigma}\eta) + h.c.,
    \label{eq: NR Lagrangian}
\end{align}
where $\eta\,(\eta^\dagger)$ is the operator annihilating (creating) a non-relativistic electron $e^-$, while $\xi\,(\xi^\dagger)$ creates (annihilates) its antiparticle $e^+$, namely, a non-relativistic positron. After introducing the two-body fields describing the $e^- e^+$ systems, $\Phi_1^{(\dagger)}(\vec{r},x)$ and $\Phi_3^{(\dagger)}(\vec{r},x)$, which couple to the currents $\bar{\nu}_a \gamma^0 N$ and $\bar{\nu}_a \vec{\gamma} N$, and integrating out the component fields $\eta$ and $\xi$ in the above Lagrangian, we obtain the so-called potential non-relativistic Lagrangian,
\begin{align}
    \mathcal{L}_{\rm \Phi_1} =
    &
    \int d^3 r\,
    \Phi_1^\dagger(\vec{r},x)
    \left[
        i\partial_{t} + \frac{\nabla^2_x}{4m_e}+\frac{\nabla^2_r}{m_e}+\frac{\alpha}{\abs{\vec{r}}}
    \right]\,\Phi_1(\vec{r},x)
    \nonumber \\
    &
    -G_F \Theta_{11} \left( \bar{\nu}_1 \gamma^0 N + \bar{N} \gamma^0 \nu_1 \right) \left( e^{-2im_e x^0} \Phi_1(\vec{0},x) + e^{2im_e x^0} \Phi_1^\dagger (\vec{0},x) \right) \nonumber \\
    &
    +G_F \Theta_{21} \left( \bar{\nu}_2 \gamma^0 N + \bar{N} \gamma^0 \nu_1 \right) \left( e^{-2im_e x^0} \Phi_1(\vec{0},x) + e^{2im_e x^0} \Phi_1^\dagger (\vec{0},x) \right) \nonumber \\
    &
    +G_F \Theta_{31} \left( \bar{\nu}_3 \gamma^0 N + \bar{N} \gamma^0 \nu_1 \right) \left( e^{-2im_e x^0} \Phi_1(\vec{0},x) + e^{2im_e x^0} \Phi_1^\dagger (\vec{0},x) \right),
    \label{eq: NR_Phi1}
\end{align}
for $N \to (e^-e^+)_{{}^1S_0} \nu$ at the leading order in $\beta$ and $\alpha$. Here, $\Phi_1(\vec{r},x)$ and $\Phi_1^\dagger(\vec{r},x)$ annihilates and creates the non-relativistic $e^- e^+$ two-body state with total spin zero, respectively. The arguments $\vec{r}$ and $x = (t, \vec{x})$ denote the relative and the center-of-mass coordinates of this two-body system. The NR Lagrangian for another process, $N \to (e^-e^+)_{{}^3S_1} \nu$, is obtained as
\begin{align}
    \mathcal{L}_{\vec{\Phi}_3} =
    &
    \int d^3 r\,
    \vec{\Phi}_3^\dagger(\vec{r},x)
    \left[
        i\partial_{t} + \frac{\nabla^2_x}{4m_e}+\frac{\nabla^2_r}{m_e}+\frac{\alpha}{\abs{\vec{r}}}
    \right]\,\vec{\Phi}_3(\vec{r},x)
    \nonumber \\
    &
    -G_F \Theta_{11} (1 + 4 s_W^2) \left( \bar{\nu}_1 \vec{\gamma} N + \bar{N} \vec{\gamma} \nu_1 \right) \left( e^{-2im_e x^0} \vec{\Phi}_3(\vec{0},x) + e^{2im_e x^0} \vec{\Phi}_3^\dagger (\vec{0},x) \right) \nonumber \\
    &
    +G_F \Theta_{21} (1 - 4 s_W^2) \left( \bar{\nu}_2 \vec{\gamma} N + \bar{N} \vec{\gamma} \nu_1 \right) \left( e^{-2im_e x^0} \vec{\Phi}_3(\vec{0},x) + e^{2im_e x^0} \vec{\Phi}_3^\dagger (\vec{0},x) \right) \nonumber \\
    &
    +G_F \Theta_{31} (1 - 4 s_W^2) \left( \bar{\nu}_3 \vec{\gamma} N + \bar{N} \vec{\gamma} \nu_1 \right) \left( e^{-2im_e x^0} \vec{\Phi}_3(\vec{0},x) + e^{2im_e x^0} \vec{\Phi}_3^\dagger (\vec{0},x) \right),
    \label{eq: NR_Phi3}
\end{align}
where the fields $\vec{\Phi}_3(\vec{r},x)$ and $\vec{\Phi}_3^\dagger(\vec{r},x)$ annihilates and creates the $e^-e^+$ system with total spin one (i.e., $2S+1=3$), respectively. The detailed derivation can be found in Ref.\,\cite{Matsumoto:2022ojl}.

It is convenient to expand the fields, $\Phi_1(\vec{r},x)$ and $\vec{\Phi}_3(\vec{r},x)$, using the solutions of the Schr\"{o}dinger equation (i.e., the equation of motion for the fields describing the relative motion between an electron and a positron) to calculate the decay widths of $N \to (e^-e^+)_{{}^1S_0} \nu$, $(e^-e^+)_{{}^3S_1} \nu$, including the Sommerfeld effect. The expansion of the two fields is given by
\begin{align}
    \Phi_1(\vec{r},x) =&
    \sum_{\ell,\,m} \int_0^\infty \frac{dk}{2\pi}\,C_{k \ell m}(x)\,\psi_{k \ell m}(\vec{r})
    + \cdots,
    \nonumber \\
    \vec{\Phi}_3(\vec{r},x) =&
    \sum_{\ell,\,m} \int_0^\infty \frac{dk}{2\pi}\,\vec{D}_{k \ell m}(x)\,\psi_{k \ell m}(\vec{r})
    + \cdots,
    \label{eq: expansion}
\end{align}
where $\psi_{k \ell m}(\vec{r})$ is the solution describing a continuum state of the two-body $e^-e^+$ system with wave number $k = (m_e E)^{1/2}$ and azimuthal and magnetic quantum numbers $\ell$ and $m$, respectively, where $E$ is the internal (kinetic) energy of the two-body system. We omit the contributions to the fields $\Phi_1(\vec{r},x)$ and $\vec{\Phi}_3(\vec{r},x)$ from the solutions describing bound states of the two-body system, as we are interested in the region $E \geq 0$. The coefficients $C_{k \ell m}(x)$ and $\vec{D}_{k \ell m}(x)$ are the fields that annihilate the continuum states with total spin zero and one. Here, the solution $\psi_{k \ell m}(\vec{r})$ is required to satisfy a specific normalization condition,
\begin{align}
    \int d^3r\,\psi^\dagger_{k' \ell' m'}(x)\,\psi_{k \ell m}(x) =
    (2\pi)\,\delta(k - k')\,\delta_{\ell \ell'} \delta_{m m'},
\end{align}
to ensure that the kinetic terms of the fields $C_{k \ell m}(x)$ and $\vec{D}_{k \ell m}(x)$ are canonically normalized. Since the Schr\"{o}dinger equation has a Coulomb potential, its solution is given by\,\cite{landau},
\begin{align}
    \psi_{k \ell m}(\vec{r}) =
    \frac{\Gamma[1 + \ell + i\alpha m_e/(2k)]}{(2\ell + 1)!r}
    \exp\left(\frac{\pi \alpha m_e}{4k}\right)
    M(i\alpha m_e/k, \ell + 1/2, -2ikr)
    Y_{\ell m}(\theta,\varphi),
    \label{eq: Coulomb_Continuum}
\end{align}
where $Y_{\ell m}(\theta, \varphi)$, $M(a,b,c)$, and $\Gamma(x)$ denote the spherical harmonic function, the Whittaker function (of the first kind), and the Gamma function, respectively. Substituting this solution into the Lagrangian for the two-body system with total spin zero in Eq.\,(\ref{eq: NR_Phi1}) gives
\begin{align}
    \mathcal{L}_{\Phi_1} =
    \label{eq: pNR_Phi1}
    &
    \int \frac{dk}{2\pi} C^\dagger_{k00} \left[ i\partial_t + \frac{\nabla^2}{4m_e} - \frac{k^2}{m_e} \right] C_{k00}
    \\
    &
    -G_F \Theta_{11} \left( \bar{\nu}_1 \gamma^0 N + \bar{N} \gamma^0 \nu_1 \right) \left[ e^{2im_e x^0} \int \frac{dk}{2\pi} \left(\frac{\alpha m_e k}{1 - e^{-\pi\alpha m_e/k}} \right)^{1/2} C^\dagger_{k00} + h.c. \right] \nonumber \\
    &
    +G_F \Theta_{21} \left( \bar{\nu}_2 \gamma^0 N + \bar{N} \gamma^0 \nu_2 \right) \left[ e^{2im_e x^0} \int \frac{dk}{2\pi} \left(\frac{\alpha m_e k}{1 - e^{-\pi\alpha m_e/k}} \right)^{1/2} C^\dagger_{k00} + h.c. \right] \nonumber \\
    &
    +G_F \Theta_{31} \left( \bar{\nu}_3 \gamma^0 N + \bar{N} \gamma^0 \nu_3 \right) \left[ e^{2im_e x^0} \int \frac{dk}{2\pi} \left(\frac{\alpha m_e k}{1 - e^{-\pi\alpha m_e/k}} \right)^{1/2} C^\dagger_{k00} + h.c. \right],
    \nonumber
\end{align}
where we have omitted writing the fields with the quantum number $\ell \neq 0$, as their wave functions vanish at the origin, $\vec{r} = 0$, and they do not couple directly to the neutrino current, $\bar{\nu}_i \gamma^0 N + \bar{N} \gamma^0 \nu_i$, at the leading order of the NR expansion. Substituting the solution\,(\ref{eq: Coulomb_Continuum}) into the Lagrangian for the two-body system with total spin one, given in Eq.\,(\ref{eq: NR_Phi3}), gives
\begin{align}
    \mathcal{L}_{\Phi_3} =
    \label{eq: pNR_Phi3}
    &
    \int \frac{dk}{2\pi} \vec{D}^\dagger_{k00} \left[ i\partial_t + \frac{\nabla^2}{4m_e} - \frac{k^2}{m_e} \right] \vec{D}_{k00}
    \\
    &
    -G_F \Theta_{11} (1 + 4 s_W^2) \left( \bar{\nu}_1 \vec{\gamma} N + \bar{N} \vec{\gamma} \nu_1 \right) \left[ e^{2im_e x^0} \int \frac{dk}{2\pi} \left(\frac{\alpha m_e k}{1 - e^{-\pi\alpha m_e/k}} \right)^{1/2} \vec{D}^\dagger_{k00} + h.c. \right] \nonumber \\
    &
    +G_F \Theta_{21} (1 - 4 s_W^2) \left( \bar{\nu}_2 \vec{\gamma} N + \bar{N} \vec{\gamma} \nu_2 \right) \left[ e^{2im_e x^0} \int \frac{dk}{2\pi} \left(\frac{\alpha m_e k}{1 - e^{-\pi\alpha m_e/k}} \right)^{1/2} \vec{D}^\dagger_{k00} + h.c. \right] \nonumber \\
    &
    +G_F \Theta_{31} (1 - 4 s_W^2) \left( \bar{\nu}_3 \vec{\gamma} N + \bar{N} \vec{\gamma} \nu_3 \right) \left[ e^{2im_e x^0} \int \frac{dk}{2\pi} \left(\frac{\alpha m_e k}{1 - e^{-\pi\alpha m_e/k}} \right)^{1/2} \vec{D}^\dagger_{k00} + h.c. \right],
    \nonumber
\end{align}
where, as in the case of $\Phi_1$, we have omitted the fields with $\ell \neq 0$, since they do not couple directly to the neutrino current, $\bar{\nu}_1 \vec{\gamma} N + \bar{N} \vec{\gamma} \nu_1$, at leading order in the NR expansion.

Using these potential NR Lagrangians, the partial decay widths of $N \to (e^-e^+)_{{}^1S_0} \nu$ and $N \to (e^-e^+)_{{}^3S_1} \nu$ at the $e^- e^+$ threshold region are obtained through the LSZ reduction formula.
In the former process, $N \to (e^-e^+)_{{}^1S_0} \nu$, an asymptotic field corresponding to $C_{k00}(x)$ in the Lagrangian\,(\ref{eq: pNR_Phi1}), which describes the $e^- e^+$ two-body state with total spin zero, satisfies the equation $[i\partial_t + \nabla^2/(4 m_e) - k^2/m_e] C^{(\rm as)}_{k00}(x) = 0$; it admits the following solution:
\begin{align}
    C_{k00}^{({\rm as})}(x) =
    -\int \frac{d^3p}{\sqrt{(2\pi)^3 2E_{p,k}}}
    \frac{1}{2\pi}\,A^{({\rm as})}_k(\vec{p})\,
    e^{-iE_{p,k} x^0 + i\vec{p}\cdot\vec{x}}.
    \label{eq: LSZ1}
\end{align}
Here, $E_{p,k} = \vec{p}^{\,2}/(4m_e) + k^2/m_e$. According to the equal-time commutation relation for the canonical variable $C^{({\rm as})}_{k00}(x)$, $[C^{({\rm as})}_{k00}(t, \vec{x}),\,\Pi^{({\rm as})}_{k'00}(t, \vec{y})] = i\delta(\vec{x} - \vec{y}) \delta(k-k')$, with the canonical conjugate $\Pi^{({\rm as})}_{k00} = iC_{k00}^{({\rm as})\,\dagger}/(2\pi)$, the operator $A^{({\rm as})}_k(\vec{p})$ satisfies the commutation relation:
\begin{align}
    [A^{({\rm as})}_k(\vec{p}), \,A^{({\rm as})\,\dagger}_{k'}(\vec{p}')] =
    (2\pi)^3\,(2E_{p,k})\,\delta(\vec{p} - \vec{p}')\,\delta(k - k').
    \label{eq: commutation relation}
\end{align}
Using the solution\,(\ref{eq: LSZ1}), the operators $A^{({\rm as})}_k(\vec{p})$ and $A^{({\rm as})\,\dagger}_k(\vec{p})$ are described in terms of the annihilation and creation operators for the $e^-e^+$ state, $C^{({\rm as})}_{k00}(x)$ and $C^{({\rm as})\,\dagger}_{k00}(x)$, as follows:
\begin{align}
    A_k^{({\rm as})\,(\dagger)}(\vec{p}) 
    = \int d^3x\,
    f_{\vec{p},k}^{(*)}(x)\,
    C^{({\rm as})\,(\dagger)}_{k00}(x)
    \quad {\rm with} \quad
    f_{\vec{p}, k}(x) =
    -\sqrt{\frac{E_{p,k}}{\pi}} e^{-iE_{p,k} x^0 + i\vec{p}\cdot\vec{x}}.
\end{align}
By defining the $e^-e^+$ two-body state as $\ket{e^-e^+(\vec{p}, k), {\rm as}} = A^{({\rm as})\,\dagger}_k(\vec{p}) \ket{0}$, which is normalized by $\braket{e^-e^+(\vec{p}, k), {\rm as}| e^-e^+(\vec{p}', k'), {\rm as}} = (2\pi)^3\,2E_{p,k} \,\delta(\vec{p} - \vec{p}') \delta(k - k')$ because of Eq.\,(\ref{eq: commutation relation}), the amplitude for the final state $\ket{e^-e^+(\vec{p}, k), {\rm out}}$ is obtained via the LSZ reduction formula as
\begin{align}
    \braket{e^-e^+(\vec{p}, k), {\rm out}|{\rm in}} =
    -i \int d^4x\,
    f_{\vec{p},k}^*(x)
    \left(
        i \partial_0 + \frac{\vec{\nabla}^2}{4m_e} - \frac{k^2}{m_e}
    \right)
    \bra{0} T[C_{k00}(x) \cdots] \ket{0},
\end{align}
where ``$\cdots$'' represents the operators responsible for generating the initial state $\ket{\rm in}$. The transition matrix element associated with the process $N \to (e^-e^+)_{{}^1S_0} \nu_a$ is then given by
{\small
\begin{align}
    \braket{e^-e^+(\vec{p},k), \nu_a(\vec{k}')|N(\vec{p}')} =
    (-i)
    &\int d^4x\,
    f_{\vec{p},k}^*(x)
    \left(
        i \partial_0 + \frac{\vec{\nabla}^2}{4m_e} - \frac{k^2}{m_e}
    \right)
    \\
    \times
    (-i)
    &\int d^4x'\, \bar{u}_{\nu_a}(\vec{k}')\,e^{ik^\prime \cdot x^\prime}
    \left(
        i\gamma_\mu\partial_{x'}^\mu
    \right)
    \nonumber \\
    \times
    (+i)
    &\int d^4y\,
    \bra{0} T[C_{k00}(x) \nu_i(x') \bar{N}(y)] \ket{0}
    \left(
        i\gamma_\nu\overleftarrow{\partial_y^\nu} + m_N
    \right)
    u_N(\vec{p}')\,e^{-ip^\prime \cdot y},
    \nonumber
\end{align}
}\noindent
where the four-momentum vectors $k'$ and $p'$ are given by $k' = (|\vec{k}'|, \vec{k})^T$ and $p' = (E_{p'}, \vec{p}')^T$, with $E_{p'} = \sqrt{|\vec{p}'|^2 + m_N^2}$.  
The wave functions of $\nu_a$ and $N$ are denoted by $u_{\nu_a}(\vec{k}')$ and $u_N(\vec{p}')$, respectively. Since the invariant amplitude appears in the transition matrix element as
\begin{align}
    \braket{e^-e^+(\vec{p},k), \nu_a(\vec{k}') | N(\vec{p}')} =
    i (2\pi)^4\, \delta^{(4)}(p' - p - k')\,
    \mathcal{M}[N \rightarrow (e^-e^+)_{{}^{1}S_0} + \nu_a],
\end{align}
where $p = (E_{p,k}, \vec{p})^T$, it is obtained at leading order in the interaction given in Eq.\,(\ref{eq: pNR_Phi1}) as
\begin{align}
    {\cal M}[N \rightarrow (e^-e^+)_{{}^{1}S_0} + \nu_a] =
    \pm [(2\pi)^3\, 2E_{p,k}]^{1/2}
    \frac{G_F\,\Theta_{a1}}{4 \pi^2}
    (\bar{u}_\nu \gamma^0 P_L\,u_N)
    \left(
        \frac{\alpha m_e k}{1 - e^{-\pi \alpha m_e /k}}
    \right)^{1/2}.
    \label{eq: Sommerfeld amplitude eenu0}
\end{align}
The sign at the beginning of the right-hand side of the equation is positive when the final state is $\nu_1$ (i.e., $a = 1$), and negative when the final state is $\nu_2$ or $\nu_3$ (i.e., $a = 2$ or $3$).

\begin{figure}[t]
    \centering
    \includegraphics[keepaspectratio, scale=0.6]{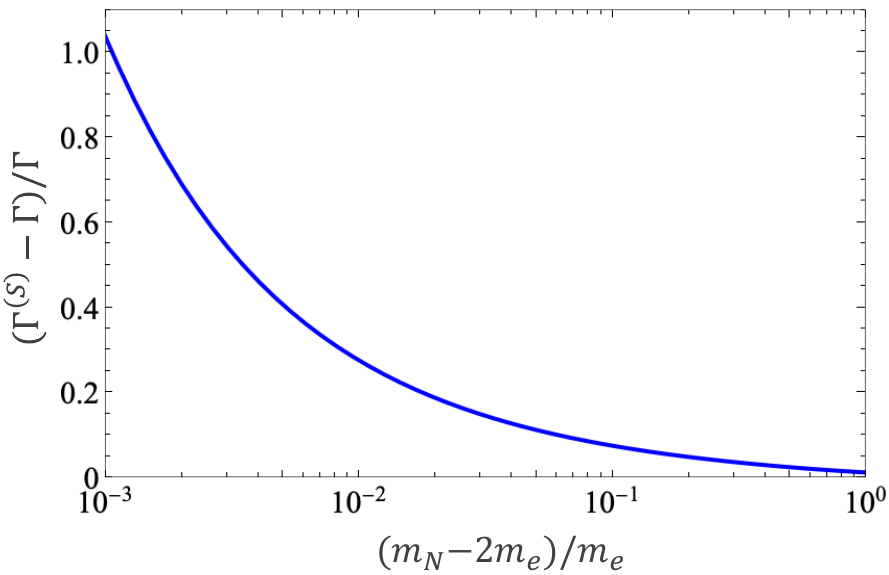}
    \caption{\small \sl
    The relative difference between the decay widths with and without the Sommerfeld correction, $(\Gamma^{(S)}[N \to e^- e^+ \nu] - \Gamma[N \to e^- e^+ \nu]) / \Gamma[N \to e^- e^+ \nu]$, is shown as a function of $m_N - 2 m_e$.}
    \label{fig: eenu}
\end{figure}

Then, the partial decay width of $N \rightarrow (e^-e^+)_{{}^{1}S_0}$, is obtained as via the decay formula as
{\small
\begin{align}
    \Gamma[N \rightarrow (e^-e^+)_{{}^{1}S_0} + \nu_a]
    &=
    \frac{1}{2 m_N}
    \int
    \overline{\sum}
    \left|
        {\cal M}[N \rightarrow (e^-e^+)_{{}^{1}S_0} + \nu_a]
    \right|^2\,
    d\Phi
    \\
    &=
    \frac{1}{2 m_N}
    \int dk \int \frac{d^3 p}{(2\pi)^3\,2E_p} \int\frac{d^3 k'}{(2 \pi)^3\,2 |\vec{k}'|} (2 \pi)^4 \delta(p' - p - k')
    |{\cal M}[\cdots]|^2,
    \nonumber
\end{align}
}\noindent
where we take the rest frame of the sterile neutrino dark matter, $p' = (m_N, \vec{0})^T$. The symbol $\overline{\sum}$ denotes the summation over the spin states of the final-state neutrino, while averaging over the spin states of the dark matter in the initial state. Calculating the squared amplitude and performing the phase space integration over $k$, $\vec{p}$, and $\vec{k}$ yields the result as follows:
\begin{align}
    \Gamma_{\rm NR}^{(S)}[N \to (e^-e^+)_{{}^1S_0} \nu]
    =
    \frac{G_F^2 m_N^5}{32\pi^3}
    \left[
        \sum_a \Theta_{a1}^2
    \right]\,I^{(s)}\left(\frac{m_N}{m_e}\right),
    \label{eq: Sommerfeld eenu 0}
\end{align}
which is almost the same as that in Eq.\,(\ref{eq: NR eenu 0}), except that the function $I^{(s)}(x)$ is now given by
\begin{align}
    I^{(s)}(x) =
    \int_0^{x-2} dy\,
    \frac{\pi\alpha/y^{1/2}}{1 - e^{-\pi \alpha/y^{1/2}}}
    \frac{y^{1/2} (\sqrt{x - 1 - y} - 1)^2}{2 \sqrt{x - 1 - y}}.
\end{align}
The invariant amplitude for the other decay process, $N \to (e^- e^+)_{{}^{3}S_1} + \nu_a$, is obtained in the same manner as for the decay process $N \to (e^- e^+)_{{}^1S_0} + \nu$, and is given as follows:
{\small
\begin{align}
    \mathcal{M}[N \rightarrow (e^- e^+)_{{}^{3}S_1} + \nu_a] =
    [(2\pi)^3\,2E_{p,k}]^{1/2}
    \frac{G_F \Theta_{a1}}{4 \pi^2} (4s_W^2 \pm 1)
    ( \bar{u}_{\nu_a} \vec{\epsilon} \cdot\vec{\gamma} P_L u_N ) \left( \frac{\alpha m_e k}{1 - e^{-\pi \alpha m_e /k}} \right)^{1/2},
\end{align}
}\noindent
where the sign ``$\pm$'' appearing in the above formula is taken to be positive when the final-state neutrino is $\nu_a = \nu_1$, and negative when $\nu_a = \nu_2$ or $\nu_3$, as in Eq.\,(\ref{eq: Sommerfeld amplitude eenu0}). The polarization vector of the two-body state $(e^- e^+)_{{}^{3}S_1}$ is denoted by $\vec{\epsilon}$. Squaring the above invariant amplitude and performing the final-state phase space integration yields the partial decay width as
\begin{align}
    &\Gamma_{\rm NR}^{(S)}[N \rightarrow (e^-e^+)_{{}^3S_1} \nu]
    =
    \frac{3 G_F^2 m_N^5}{32\pi^2}
    \left[
        \sum_a \Theta_{a1}^2 \left(4s_W^2 + 2\delta_{a1} -1\right)^2
    \right]\,I^{(S)}\left(\frac{m_N}{m_e}\right).
    \label{eq: Sommerfeld eenu 1}
\end{align}
Consequently, the correction to the decay width due to the Sommerfeld effect is given by $\Delta \Gamma[N \to (e^-e^+)_{{}^1S_0, {}^3S_1} \nu] = \Gamma^{(S)}[N \to (e^-e^+)_{{}^1S_0, {}^3S_1} \nu] - \Gamma_{\rm NR}[N \to (e^-e^+)_{{}^1S_0, {}^3S_1} \nu]$. Using Eqs.\,(\ref{eq: eenu}, \ref{eq: NR eenu 0}, \ref{eq: NR eenu 1}, \ref{eq: Sommerfeld eenu 0}, \ref{eq: Sommerfeld eenu 1}), the total decay width incorporating this effect can be computed, as summarized in Eqs.\,(\ref{eq: Sommerfeld eenu}, \ref{eq: Decay with Sommerfeld}) in the main text. The comparison between the decay widths with and without the Sommerfeld correction, namely $\Gamma^{(S)}[N \to e^- e^+ \nu]$ and $\Gamma[N \to e^- e^+ \nu]$, is shown in Fig.\,\ref{fig: eenu} as a function of $(m_N - 2 m_e)$. The correction enhances the decay width by approximately ${\cal O}(10)\,\%$ when $m_N$ is degenerate with $2m_e$ at the ${\cal O}(1)\,\%$ level, and by ${\cal O}(100)\,\%$ when the degeneracy reaches the ${\cal O}(0.1)\,\%$ level, as noted in the main text.

Using the potential NR Lagrangians\,(\ref{eq: pNR_Phi1},\,\ref{eq: pNR_Phi3}), the differential decay width of the sterile neutrino DM into the final state $e^- e^+ \nu$, given by $d\Gamma[N \rightarrow e^- e^+ \nu]/dE_{e^\pm}$, which is used to evaluate the cosmological constraint in Eq.\,(\ref{eq: differential decay width}), can also be obtained as follows:
{\small
\begin{align}
    \frac{d\Gamma}{dE_{e^\pm}}[N \rightarrow e^- e^+ \nu] =
    &
    \frac{d\Gamma^{(0)}}{dE_{e^\pm}}[N \rightarrow e^- e^+ \nu]
    +\frac{d(\Delta\Gamma)}{dE_{e^\pm}}[N \rightarrow (e^- e^+)_{^1S_0} \nu]
    +\frac{d(\Delta\Gamma)}{dE_{e^\pm}}[N \rightarrow (e^- e^+)_{^3S_1} \nu],
    \nonumber \\
    \frac{d\Gamma^{(0)}}{dE_{e^\pm}}[N \rightarrow e^- e^+ \nu] =
    &
    \frac{G_F^2 m_N^4}{96 \pi^3}\sum_a \Theta_{a1}^2
    \left\{ 
        \frac{(1 - 2z)^2 (z^2 - x_e^2)^{1/2}}
        {(1 - 2z + x_e^2)^3}
        \left[
            (4s_W^2 + 2\delta_{a1} -1)^2 g_1(z) + g_2(z)
        \right]
    \right\},
    \nonumber \\
    g_1(z) =
    &
    32z^3 - 2z^2 (12x_e^2 + 17) + z (6x_e^4 + 7x_e^2 + 9) + 4x_e^2,
    \nonumber \\
    g_2(z) =
    &
    32z^3 - 2z^2 (24 x_e^2 + 17) + z (18x_e^4 + 43x_e^2 + 9) - 8x_e^2 - 12x_e^4,
\end{align}
} \noindent
where the correction from the Sommerfeld effect, $d(\Delta\Gamma)[N \to (e^- e^+)_{^1S_0} \nu]/dE_{e^\pm}$, is
{\small
\begin{equation}
    \frac{d(\Delta\Gamma)}{dE_{e^\pm}}[N \rightarrow (e^- e^+)_{^1S_0} \nu] =
    \frac{G_F^2 m_N^4}{16 \pi^3}
    \left(\sum_a \Theta_{a1}^2 \right)
    \left(
        \frac{\pi\alpha/\sqrt{y}}{1 - e^{-\pi \alpha/\sqrt{y}}} -1
    \right)
    \sqrt{y} (\sqrt{x_N - 1 - y} - 1)^2,
\end{equation}
}\noindent
with the parameter $y$ being 
$y = (x_N + 2) E_{e^\pm}/m_e - x_N^2/4 - E_{e^\pm}^2/m_e^2 - 2$. Here, $x_N = m_N/m_e$. The correction to the decay into the ${}^3S_1$ channel, $d(\Delta\Gamma)[N \to (e^- e^+)_{^3S_1} \nu]/dE_{e^\pm}$, is obtained by replacing $\Theta_{a1}^2$ with $\Theta_{a1}^2 (4s_W^2 + 2\delta_{a1} -1)^2$, and multiplying the result by a factor of three.

%Sterile neutrino $(m_N, \vec{0})$,
%SM neutrino $(P,\vec{P})$,
%electron-positron pair $(2m_e + k^2/m_e + P^2/4m_e, \vec{P})$.
%Electron energy:
%\begin{align}
%    E_{e^-} = m_e + \frac{(\vec{k} + \vec{P}/2)^2}{2 m_e} = m_e + \frac{k^2}{2 m_e} + \frac{P^2}{8 m_e} + \frac{\vec{k} \cdot \vec{P}}{2 m_e}
%\end{align}
%Take the average over the angle (assumption):
%\begin{align}
%    \braket{\vec{k} \cdot \vec{P}} = \frac{\int_{S^2} d\Omega \vec{k} \cdot \vec{P}}{\int_{S^2} d\Omega} = 0
%\end{align}
%Then
%\begin{align}
%    \braket{E_{e^-}} = m_e + \frac{k^2}{2 m_e} + \frac{P^2}{8 m_e}
%\end{align}
%Energy conservation:
%\begin{align}
%    m_N &= 2m_e + \frac{k^2}{m_e} + \frac{P^2}{4 m_e} + P \nonumber \\
%    \therefore P &= -2 m_e + 2\sqrt{m_N m_e - m_e^2 -k^2} \nonumber \\
%    \therefore \braket{E_{E^-}} &= m_e + \frac{m_N}{2} - \sqrt{m_N m_e - m_e^2 -k^2}
%\end{align}
%With $x = m_N/m_e $ and $y = k^2/m_e^2$.
%\begin{align}
%    \frac{d}{d\braket{E_{e^-}}} = \frac{2 \sqrt{x -1 -y}}{m_e} \frac{d}{dy}
%\end{align}

%%%%%%%%%%%%%%%%%%%%%%%%%%%%%%%%%%%%
%%%%%%%%%%% Bibliography %%%%%%%%%%%
%%%%%%%%%%%%%%%%%%%%%%%%%%%%%%%%%%%%
\bibliographystyle{unsrt}
\bibliography{references}

\end{document}